\documentclass[twocolumn,showpacs,preprintnumbers,amsmath,amssymb]{revtex4}

\usepackage{color}

\usepackage{graphicx}
\usepackage{graphicx,epstopdf}
\usepackage{dcolumn}
\usepackage{bm}

\begin{document}


\title{Chimera states in two-dimensional networks of locally coupled oscillators}
\author{Srilena Kundu$^1$}
\author{Soumen Majhi$^1$}
  \author{Bidesh K. Bera$^1$}
  \author{Dibakar Ghosh$^1$}
\email{diba.ghosh@gmail.com}
\author{M. Lakshmanan$^2$}
\affiliation{$^1$Physics and Applied Mathematics Unit, Indian Statistical Institute, 203 B. T. Road, Kolkata-700108, India\\
	$^2$Centre for Nonlinear Dynamics, School of Physics, Bharathidasan University, Tiruchirapalli-620024, India}

\date{\today}

\begin{abstract} 
Chimera state is defined as a mixed type of collective state in which synchronized and desynchronized subpopulations of a network of coupled oscillators coexist and the appearance of such anomalous behavior has strong connection to diverse neuronal developments.  Most of the previous studies on chimera states are not extensively done in two-dimensional ensembles of coupled oscillators by taking neuronal systems with nonlinear coupling function into account while such ensembles of oscillators are more realistic from a neurobiological point of view. In this paper, we report the emergence and existence of chimera states by considering locally coupled two-dimensional networks of identical oscillators where each node is interacting through nonlinear coupling function. This is in contrast with the existence of chimera states in two-dimensional nonlocally coupled oscillators with rectangular kernel in the coupling function. We find that the presence of nonlinearity in the coupling function plays a key role to produce chimera states in two-dimensional locally coupled oscillators. We analytically verify explicitly in the case of a network of coupled Stuart - Landau oscillators in two dimensions that the obtained results using Ott-Antonsen approach and our analytical finding very well matches with the numerical results. Next, we consider another type of important nonlinear coupling function which  exists in neuronal systems, namely chemical synaptic function, through which the nearest-neighbor (locally coupled) neurons interact with each other. It is shown that such synaptic interacting function promotes the emergence of chimera states in two-dimensional lattices of locally coupled neuronal oscillators. In numerical simulations, we consider two paradigmatic neuronal oscillators, namely Hindmarsh-Rose neuron model and Rulkov map for each node which exhibit bursting dynamics. By associating various spatio-temporal behaviors and snapshots at particular times, we study the chimera states in detail  over a large range of coupling parameter. The existence of chimera states is confirmed by instantaneous angular frequency, order parameter and strength of incoherence.
\end{abstract}


\maketitle


\section{Introduction}
One of the most complex systems in the real world is the human brain and understanding the interaction between the neurons through the synapses is one of the most challenging issues. In the nervous system, synapses are functional connections that permit a neuron to pass signals to other neurons and there are essentially two different types of synapses, namely,  electrical synapse and chemical synapse. Through a chemical synapse, information passes chemically between two neurons in the form of neurotransmitter molecules, whereas an electrical synapse is a gap junction that has channel proteins connecting the two neurons, so the electrical signal can move straight over the synapse. Chemical synapses relay information through chemicals and they are sturdy but electrical synapses are not as efficient as the chemical synapses. We also note that the two-dimensional (2D) architecture of interaction between the neurons is quite likely in reality and so an organized study on the emerging behaviors in two-dimensional networks of coupled neurons is particularly necessary and important. 

\par During the unihemispheric slow-wave sleep \cite{uhsws1,uhsws2} in many aquatic mammals and migratory birds, half of their brain is awake while the remaining portion is in sleep. The neuronal oscillations are synchronized in the sleepy part, whereas the oscillations of the awake portion are desynchronized. This type of neuronal activity in the brain is intimately related to the Kuramoto's finding \cite{kuramoto} of coexistence of synchronization (coherence) and desynchronization (incoherence) in nonlocally coupled networks of identical phase oscillators, which was later named as {\it Chimera} by Strogatz \cite{strogatz}. The existence of such a state is also pertinent to various types of brain diseases \cite{brain_disease1,brain_disease2}, such as Parkinson's disease, Alzheimer's disease, epileptic seizures, schizophrenia, and brain tumors. As noted earlier, chimera is a peculiar type of synchronization phenomenon that comprises coherent and incoherent dynamics in coupled oscillatory networks. Appearance of chimera states is very fascinating, since it emerges in a network of symmetrically coupled identical oscillators \cite{chimera_rev,epl_rev}. Initially, chimera states were detected in nonlocally coupled phase oscillators with exponential coupling functions. Subsequently, such a new discovery has drawn the interests of many researchers, and it was revealed that chimera states also appear in coupled limit-cycle oscillators \cite{limit,limit2}, chaotic oscillators \cite{chaotic}, chaotic maps \cite{chaotic_map}, hyper chaotic time delay \cite{lakshman_measure,bs_chimera} systems, and neuronal systems \cite{hr_bera1,hr_ijbc,chimera_modular}. At the beginning, it was believed that chimera states emerge in coupled networks due to ushering of nonlocality in coupling configuration but many recent results are not restricted to this point; they uncovered that chimera states may appear in global (all-to-all) networks \cite{global1,global2,global3,global4,global5,global6,global7} and even in local (nearest-neighbor) networks \cite{laing,hr_bera1,hr_bera2,local1}. Beside these symmetric coupling topologies, emergence of chimera states is also possible in heterogeneous networks \cite{ch_hetero}, static and time varying complex networks \cite{complex_ch1,complex_ch2}, multiplex \cite{chimera_multiplex,multiplex1,multiplex2,multiplex3,multiplex4} and modular networks \cite{chimera_modular}, etc. Very recently chimera and chimeralike states were observed in two distinct groups of identical populations where each population is nonlocally \cite{two_nonlocal} and globally \cite{two_global} connected, respectively. Depending on the variations in amplitude, phase and the spatiotemporal behavior of the oscillators,  chimera states can be classified in different categories as amplitude mediated chimera \cite{amc}, amplitude chimera \cite{cd_prl}, imperfect chimera \cite{imperfect_chi}, traveling chimera \cite{travelling_chi}, imperfect traveling chimera \cite{hr_bera3}, breathing chimera \cite{breath1},  spiral wave chimera \cite{spiral_chi}, etc. 

\par In this context, systematic studies on chimera states in neuronal networks deserve special attention. Earlier works \cite{hr_bera1,chimera_modular,hr_ijbc,hr_bera2,multiplex1,chimera_multiplex,hr_bera3} on chimera states have been done using nonlocally, locally, and globally coupled neurons using electric and chemical synapses. However, in most of the previous works, only one-dimensional lattice of neurons was considered. Normally, the neurons in the brain are connected in two-dimensional grids and transferring the signals to the neighboring neurons takes place through synapses. Thus, it is important to study the different spatiotemporal behaviors in  two-dimensional grid  networks of locally connected neurons. Recently, different types of chimera states were investigated in two-dimensional \cite{2d} and three-dimensional \cite{3d} systems by considering different types of coupling functions in phase oscillators mainly.  In this context, Schmidt \textit{et al}. \cite{pre2017} reported the occurrence of different types of chimera states in a network of two dimensional lattice of neurons under nonlocal coupling. This study is mainly focused on nonlocal coupling with rectangular kernel in the coupling function, which has less neurological importance  as far as the coupling function is concerned. In this connection, neuronal oscillators in purely local coupled neurons in 2D lattice is one of the most realistic coupling schemes through which neurons are connected, which can be considered as an approximation of the acute brain slices. 

\par  In this paper, we systematically study the existence of chimera states in 2D grid of coupled oscillators where their interactions take place by means of nonlinear coupling function with the nearest neighbors only. In most of the previous studies \cite{pre2017,2D_linchi} on chimera states in 2D grid of oscillators, linear coupling function was used with nonlocal coupling topology.
Here we clearly articulate that nonlinearity present in the coupling function leads to the emergence of chimera states in two-dimensional locally coupled oscillators. We start with an ensemble of Stuart - Landau oscillators with nonlinear coupling function interacting solely via a nearest-neighbor coupling topology and show the existence of chimera state is observed therein. To clarify that the observed chimera state does not depend on the number of oscillators in the 2D grid, we have studied the continuous version of the model using  Ott-Antonsen method and our analytical finding very well matches with our numerical results. Next we consider each node of the 2D grid network by (i) Hindmarsh-Rose neuronal oscillator and (ii) Rulkov map. These two systems are more realistic since depending on the system parameters, they can produce different excitability and bursting dynamics. Using these two dynamical systems to cast the nodes in the 2D grid, we show that the network gives rise to chimera states as we tune the interaction strength $\epsilon$. Further all the recognized states, namely incoherent, chimera and coherent patterns are characterized using instantaneous angular frequency, Kuramoto order parameter and strength of incoherence.

\par The subsequent parts of this paper are organized as follows. Section II introduces the general mathematical frame of two-dimensional grid of networks. In Sec. III, the emergence of chimera states is discussed using Stuart-Landau oscillators. The phase reduction form and analytical results using Ott-Antonsen approach are discussed in Secs. IIIA and IIIB respectively. Sections IV and V devote the results on Hindmarsh-Rose and Rulkov models, respectively, and order parameter and strength of incoherence are used to distinguish different states such as incoherent, chimera and coherent states. Section VI provides conclusions of our finding.
\begin{figure}[ht]
\centerline{
\includegraphics[scale=0.35]{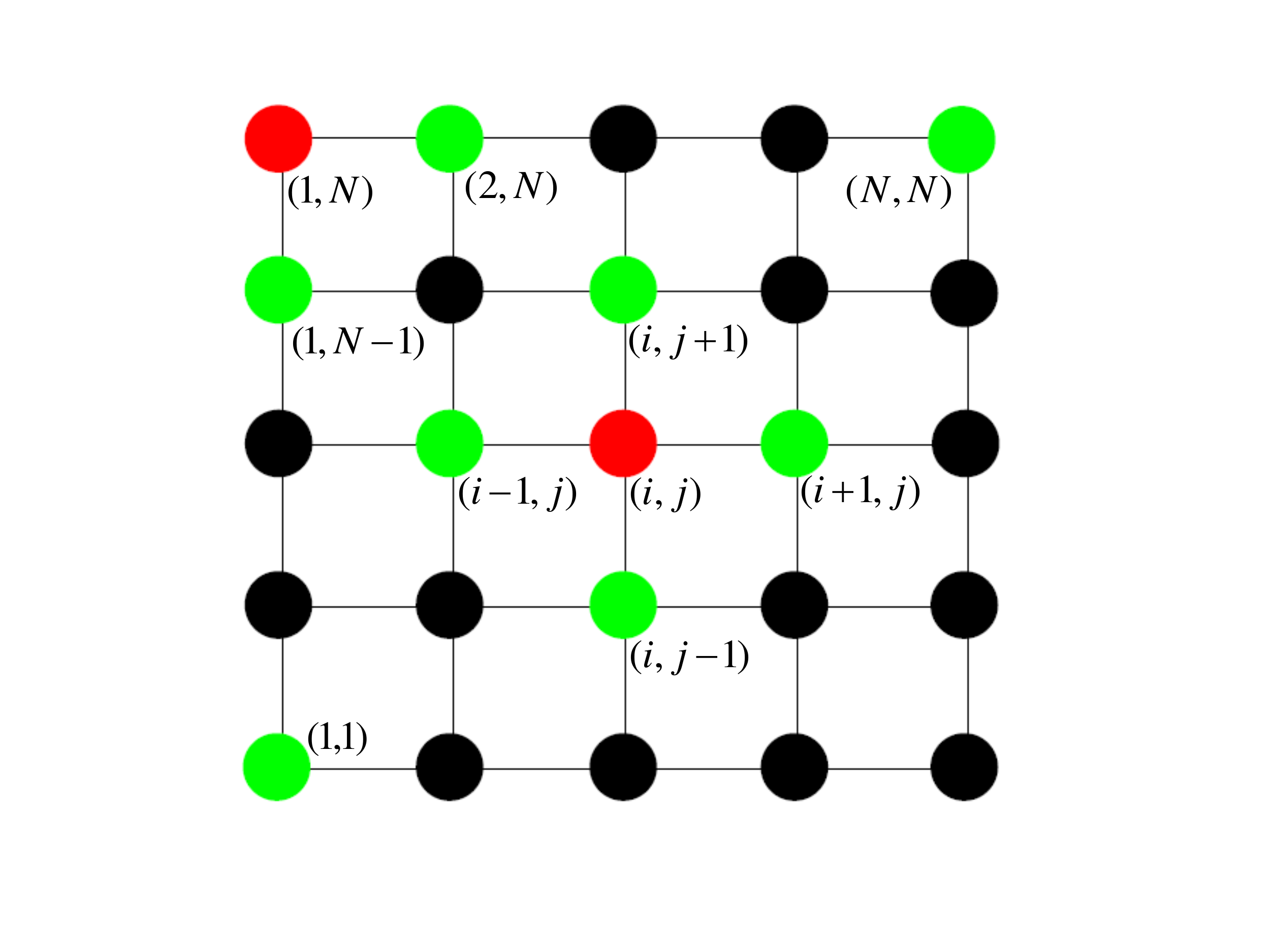}}
\caption{{ Schematic diagram of a two-dimensional grid: the $(i,j)$-th oscillator (red circle) is  connected to its four nearest-neighbor oscillators (green circle). For illustration of local coupling, we mark two nodes in the $(i,j)$-th and $(1, N)$-th position by a red circles which are connected to their nearest neighbors on four sides by green circles (discrete 2-torus). Black circles represent other nodes of the network. Each node in the grid is actually connected to nearest nodes by similar manner but not shown here for clarity of picture. Here the figure represents $N=5$. }}
\label{ 2D_Config}
\end{figure} 

\section{Mathematical form of COUPLED SYSTEMS}
  We consider a network of $N\times N$ two-dimensional grid of locally coupled oscillators as shown in Fig.~\ref{ 2D_Config}. The local dynamics of individual node of the network is given by $\dot X_{i,j}=F(X_{i,j})$, where $X_{i,j}$ represents an $l$-dimensional vector of the dynamical state variables and $F(X_{i,j})$ is the corresponding velocity field. The general mathematical equations of locally coupled systems in a 2D grid of network can be described as 
	\begin{equation} \label{eq:11}
	\begin{array}{lcl}
	\dot{X}_{i,j}=F(X_{i,j})+K\{H(X_{i,j}, X_{i-1,j})+H(X_{i,j}, X_{i+1,j})\\\\~~~~~~+H(X_{i,j}, X_{i,j-1})+H(X_{i,j}, X_{i,j+1})\},
	\end{array}
	\end{equation}
	where subscript $(i,j) (i,j=1,...,N)$ in $X_{i,j}$ and $F(X_{i,j})$ determines the position of the oscillator in the 2D coupled network. The coupling function $H:R^l \times  R^l \rightarrow R$ describes the manner by which the $(i,j)$-th oscillator is connected with its nearest-neighbor oscillators. We choose the coupling function $H$ in the form of a specific nonlinear function because using this nonlinear function we observe chimera state in 2D grid of locally coupled oscillators, and it appears that requirement of nonlinearity in the coupling function seems to be an essential criterion for the formation of chimera states under local coupling in 2D lattices. This nonlinear interaction function may develop quite naturally in some systems under consideration (e.g., nonlinear chemical synaptic functions for neuronal systems) or can be formed based on certain motives. Both of these circumstances are explained in the following for their respective occurrences.  We have also verified that if $H$ is a linear function, then the chimera state does not exist in the 2D network of locally coupled oscillators (detailed discussions are presented in the Appendix).   Here $K=(\epsilon_1, \epsilon_2,...,\epsilon_l)^T$ is the coupling matrix where $T$ denotes transpose of a matrix.  We use periodic boundary conditions in both the directions with $X_{0,j}\equiv X_{N,j}$ and $X_{i,0}\equiv X_{i,N}$, so that the array of  coupled systems has translation invariance.

	\par In the following sections, we will explore the emergence of chimera states in Eq.\eqref{eq:11} by taking three different dynamical systems, namely, (i) Stuart - Landau oscillators, (ii) Hindmarsh-Rose model, and (iii) Rulkov map with different types of coupling functions. Our main emphasis will be to identify  different collective dynamical states, including chimeras, by changing the coupling strength.

	\section{Stuart - Landau system}
		First, we consider a grid of $N\times N$ locally coupled Stuart - Landau (SL)  oscillators interacting through a nonlinear coupling function. 	The mathematical form of a single SL oscillator is given by
		\begin{equation}\label{eq:2}
		\dot{z} = (1+i\alpha)z - (1+i\beta)|z|^2z,
		\end{equation}
		where $z = x + iy$, $i=\sqrt{-1}$ and $\alpha$, $\beta$ are real parameters. Equation \eqref{eq:2} admits a generic limit cycle near a Hopf bifurcation \cite{kura_book}, where $\alpha$ is the frequency of this limit cycle.

		 The governing equations for the coupled two- dimensional network is represented by
	
	\begin{equation} \label{eq:1}
	\begin{array}{lcl}
	\dot{z}_{i,j} = (1+i\alpha)z_{i,j} - (1+i\beta)|z_{i,j}|^2z_{i,j} + \frac{\epsilon}{4}[H(z_{i-1,j})\\\\+H(z_{i+1,j}) + H(z_{i,j-1}) + H(z_{i,j+1}) - 4H(z_{i,j})],
	\end{array}
	\end{equation}
	for subscript $i,j=1,2,...,N$ with periodic boundary conditions $z_{N+1,j}=z_{1,j}, z_{i, N+1}=z_{i,1}$ and $z_{0,j}=z_{N,j}, z_{i,0}=z_{i,N}$. Here  $\epsilon$ is the coupling constant. We choose the nonlinear coupling function \cite{nonlinear_coup,moving_pre} in the form   $H(z) = \tilde{a}^2z - z|z|^2,$ where $\tilde{a}$ is real constant.

	\begin{figure}[ht]
		\centerline{
			\includegraphics[scale=0.46]{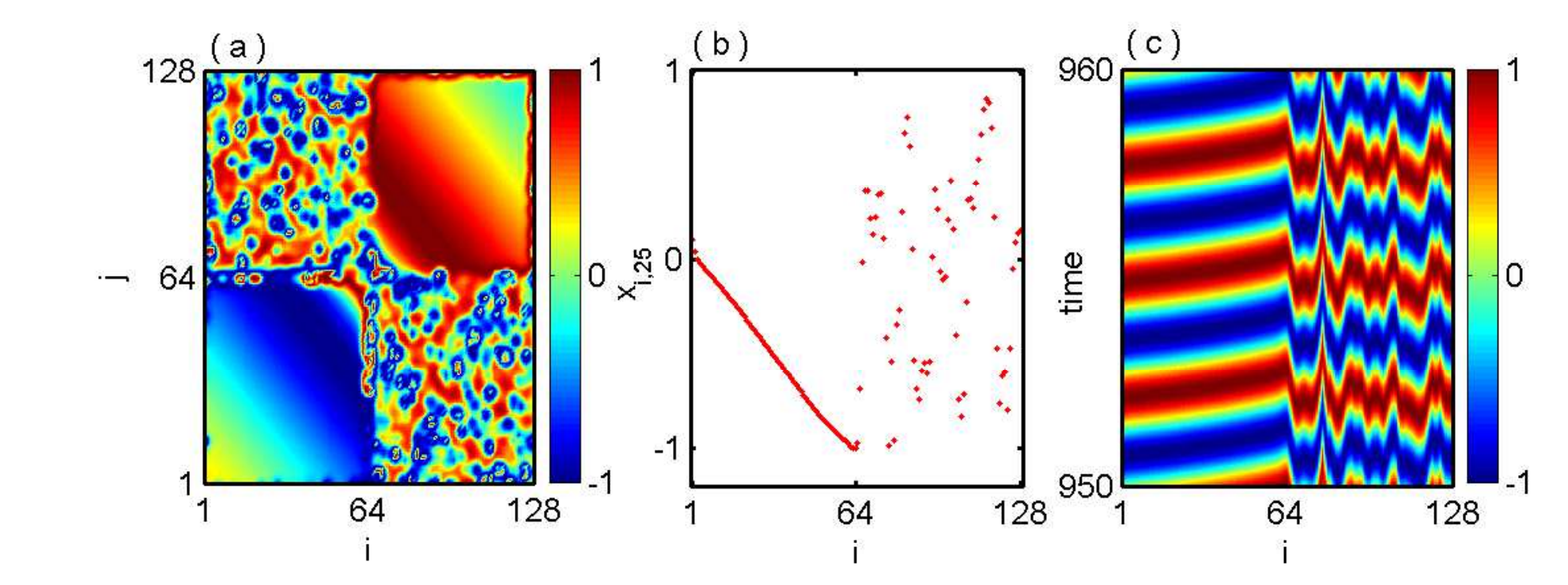}}
		\caption{ Snapshot of  (a) the state variables $x_{i,j}$ in the 2D grid, (b) $x_{i, 25}$ along the horizontal cross-section line  $j=25$ at time $t=952$, (c) space-time evolution of the state variables along this cross-section line. Here $\alpha=1.0, \beta=-1.5, \tilde a=1.02, \epsilon=0.15$.}
		\label{lsrevisefig1}
	\end{figure}
	\par  We will now first bring out the results obtained through numerical investigation and then give appropriate theoretical justification. Figure \ref{lsrevisefig1} (a) shows the snapshot of the state variables $x_{i,j}$ of the SL oscillators over the entire 2D lattice with $N=128$. For the numerical simulations, the fifth-order Runge-Kutta-Fehlberg method with time-step size of $0.01$ has been used. The initial conditions are chosen as $x_{i,j}(0)=0.001[N-(i+j)], y_{i,j}(0)=0.002[N-(i+j)], i,j=1,...,N$ with added small random fluctuations. 
	From the figure, coexistence of coherence and incoherence and consequently the chimera pattern is easily discernible. Here the oscillators having indices approximately $(1\le i \le 64)\wedge (1\le j \le 64)$ and $(65\le i \le 128)\wedge (65\le j \le 128)$ form the coherent domain while with $(1\le i \le 64)\wedge (65\le j \le 128)$ and $(65\le i \le 128)\wedge (1\le j \le 64)$ make the incoherent domain. Snapshot of the state variables $x_{i,25}$ along the horizontal cross-section $j=25$ is shown in Fig. \ref{lsrevisefig1} (b) that distinguishes coherent and incoherent groups in one-dimensional view of the lattice. Further, Fig. \ref{lsrevisefig1} (c) depicts the space-time evolution of the state variables along the same cross-section effectively signifying stationarity of the chimera state over time.
	
	\subsection{Phase reduction of coupled Stuart - Landau oscillator}
			We consider the phase reduction method \cite{nakao2015} for nonlinear oscillators to simplify Eq. \eqref{eq:1} into a set of coupled phase oscillators. The universal result of any system does not depend on the phase reduction theory, and this approach helps one to study the physical characteristic of any systems that are derived from original system more easily. 
	
 Now Eq. \eqref{eq:2} translates in Cartesian form as
	\[
	\begin{pmatrix}
	\dot{x}\\
	\dot{y}
	\end{pmatrix} 
	=
	\begin{pmatrix}
	x-\alpha y -(x - \beta y)(x^2 + y^2)\\
	\alpha x + y -(\beta x + y)(x^2 + y^2) 
	\end{pmatrix}, 
	\]	
	where $z = R\exp(i\phi)$, $R = |z| = \sqrt{x^2 + y^2}$ is the modulus and $\phi = \arctan(y/x) $ is the argument.
	
	Then $R$ and $\phi$ follows
	\begin{equation}
	\frac{dR(t)}{dt} = R - R^3, \qquad \frac{d\phi(t)}{dt} = \alpha - \beta R^2.
	\end{equation}	
	Without loss of generality, we can choose (asymptotically) $R(t) =1$ and $\phi(t) = (\alpha - \beta)t$ so that the frequency and period of the oscillation are given by 
	$\omega = \frac{d\phi}{dt} = \alpha - \beta$ and $T = \frac{2\pi}{\omega} = \frac{2\pi}{\alpha - \beta}$, respectively.
	
	The phase function of the above SL oscillator Eq.  \eqref{eq:2} is given by \cite{syn_book}
	\begin{center}
		$\Theta(z) = \Theta(R,\phi) = arg(z) - \beta \ln |z| = \phi - \beta \ln R.$
	\end{center}
	
	So the phase $\theta(t) = \Theta[z(t)]$ of the SL oscillator obeys
	\begin{center}
		$\frac{d\theta(t)}{dt} = \frac{d\phi}{dt} - \frac{\beta}{R} \frac{dR}{dt} = \alpha - \beta = \omega.$
	\end{center} 
	
	The limit cycle can be expressed as $z_0(\theta) = e^{i\theta}$ or $X_0(\theta) = (x_0(\theta), y_0(\theta)) = (\cos\theta, \sin\theta)$ as a function of the phase variable $\theta(0\leq\theta\leq 2\pi)$.
	
	The phase sensitivity function $\textbf Z(\theta) = (Z_x(\theta), Z_y(\theta))$ can be obtained by differentiating $\Theta(z)$ with respect to $x$ and $y$ as
	\begin{align*}
		\textbf Z(\theta)~=& ~(Z_x(\theta), ~ Z_y(\theta))\\
		= &~\left(\frac{\partial \theta}{\partial x}, ~ \frac{\partial \theta}{\partial y}\right)_{(x,y) = (x_0(\theta),y_0(\theta))}\\
		= &~\left(\frac{\partial \phi}{\partial x} - \frac{\beta}{R} \frac{\partial R}{\partial x}, ~ \frac{\partial \phi}{\partial y} - \frac{\beta}{R}\frac{\partial R}{\partial y}\right)_{(x,y) = (x_0(\theta),y_0(\theta))}\\
		= &~\left(-y - \beta x, ~ x - \beta y\right)_{(x,y) = (x_0(\theta),y_0(\theta))}\\
		= &~\left(-\sin \theta - \beta \cos \theta, ~\cos \theta - \beta \sin \theta\right).
	\end{align*}
	
	
	Now Eq. \eqref{eq:1} can be written in summation form as 
	\begin{equation} \label{eq:p5}
	\dot{z}_{i,j} = F(z_{i,j})+\frac{\epsilon}{4}\sum_{m=1}^{N}\sum_{n=1}^{N} A_{ijmn} \quad [H(z_{m,n})-H(z_{i,j})]
	\end{equation}
	where $A_{ijmn} $is the connectivity matrix
	 \begin{align*}
		A_{ijmn} ~ = & ~ 1, ~~\text{if}~~ m=i, n=j-1, j+1\\ & ~~~~~~\text{and} ~~n=j, m=i-1,i+1\\
		= & ~ 0, \quad \text{otherwise},
	\end{align*}
	with $F(z_{i,j})=(1+i\alpha)z_{i,j} - (1+i\beta)|z_{i,j}|^2z_{i,j}$
	and \begin{align*}
		H(z) =& ~ \tilde{a}^2z - z|z|^2\\
		=& ~ \tilde{a}^2x - x(x^2 + y^2) + i[\tilde{a}^2y - y(x^2 + y^2)]\\
		=& ~ f(x,y) + ig(x,y).
	\end{align*}
	Therefore, Eq. \eqref{eq:p5} becomes
	\begin{equation*}
		\dot{z}_{i,j} = F(z_{i,j}) + \frac{\epsilon}{4}\sum_{m=1}^{N}\sum_{n=1}^{N} A_{ijmn} ~ G(H_{i,j},H_{m,n}),
	\end{equation*}
	where \begin{align*}
		G(H_{i,j},H_{m,n})~ = & ~ [H(z_{m,n}) - H(z_{i,j})]\\
		= & ~ \begin{pmatrix}
			f_{m,n} - f_{i,j}\\
			g_{m,n} - g_{i,j}
		\end{pmatrix}.
	\end{align*}
	In terms of the phase variable $\theta$, the coupling function $G$ can be written as
	\begin{align*}
		G(\theta_{i,j},\theta_{m,n}) ~ = ~ \begin{pmatrix}
			(\tilde{a}^2 - 1)[\cos (\theta_{m,n}) - \cos (\theta_{i,j})]\\
			(\tilde{a}^2 - 1)[\sin (\theta_{m,n}) - \sin (\theta_{i,j})]
		\end{pmatrix},
	\end{align*}
	since $f_{i,j}$ = $(\tilde{a}^2-1)\cos (\theta_{i,j})$ and $g_{i,j}$ = $(\tilde{a}^2-1)\sin (\theta_{i,j})$.
	
	So the reduced phase equation for the $(i,j)$-th oscillator is given by
	\begin{equation}\label{eq:5}
	\dot{\theta}_{i,j} = \omega_{i,j} + \frac{\epsilon}{4}\sum_{m=1}^{N}\sum_{n=1}^{N} A_{ijmn} ~ \Gamma(\theta_{i,j} - \theta_{m,n}),
	\end{equation}
	where $\omega_{i,j}=\omega=\alpha-\beta$ is the natural frequency and the phase coupling function
	\begin{align*}
		\Gamma(\varphi) ~= & ~\frac{1}{2\pi}\int_{0}^{2\pi} \textbf Z(\varphi+\psi) G(\varphi+\psi,\psi) d\psi\\
		= & ~\frac{1}{2\pi}\int_{0}^{2\pi} (\tilde{a}^2 -1)(-\sin\varphi - \beta \cos \varphi + \beta) d\psi\\
		= & ~(\tilde{a}^2 -1)(-\sin\varphi - \beta \cos \varphi + \beta).
	\end{align*}
	Therefore, 
	\begin{equation*}
		\begin{array}{lcl}
			\Gamma(\theta_{i,j} - \theta_{m,n}) ~=~ (\tilde{a}^2 -1)[-\sin(\theta_{i,j} - \theta_{m,n})\\~~~~~~~~~~~~~~~~~~~~~~-\beta \cos (\theta_{i,j} - \theta_{m,n}) + \beta]	
		\end{array}	
			\end{equation*}
			This $\Gamma$ represents the effect of $(m,n)$-th oscillator on $(i,j)$-th oscillator over one period of limit cycle oscillation. The phase coupling function $\Gamma$ in Eq. ( \ref{eq:5}) depends only on the phase difference $(\theta_{i,j}-\theta_{m,n})$, which makes it easier to analyze the synchronized and desynchronized states significantly.
	Equation \eqref{eq:5} becomes
	\begin{equation*}
		\begin{array}{lcl}
			\dot{\theta}_{i,j}~=~\omega_{i,j}+\frac{\epsilon}{4}(\tilde{a}^2-1)\\\\~~~~~~~\sum_{m=1}^{N}\sum_{n=1}^{N}A_{ijmn}[-\sin(\theta_{i,j}-\theta_{m,n})-\\\\~~~~~~~~
			\beta\cos(\theta_{i,j}-\theta_{m,n})+\beta]\\\\~~~~
			=~\omega_{i,j}-\frac{\epsilon}{4}(\tilde{a}^2-1)\\\\~~~\sum_{m=1}^{N}\sum_{n=1}^{N} A_{ijmn}[\sqrt{1+\beta^2}\sin(\theta_{i,j}-\theta_{m,n}+\gamma)-\beta]\\\\~~~~
			=~\omega_{i,j}+\epsilon\beta(\tilde{a}^2-1)-\frac{\epsilon}{4}(\tilde{a}^2-1)\sqrt{1+\beta^2}\\\\~~~~~\sum_{m=1}^{N}\sum_{n=1}^{N} A_{ijmn}\sin(\theta_{i,j}-\theta_{m,n}+\gamma),
		\end{array}	
	\end{equation*}
	where $\gamma=\tan^{-1}\beta$.
	
	Finally, the phase reduced model of the two dimensionally coupled SL oscillator is given by
	\begin{equation}\label{eq:6}
	\dot{\theta}_{i,j} = \omega'_{i,j} - \lambda \sum_{m=1}^{N}\sum_{n=1}^{N} A_{ijmn} \sin(\theta_{i,j} - \theta_{m,n} + \gamma),
	\end{equation}
	where $\omega'_{i,j}$ = $\omega_{i,j} + \epsilon\beta(\tilde{a}^2-1)$, $\lambda$ = $\frac{\epsilon}{4} (\tilde{a}^2 -1) \sqrt{1+\beta^2}$ and $\gamma = \tan^{-1}\beta$.

	\subsection{Analytical results: Ott-Antonsen approach}
	
	Now we want to analytically show that the observed chimera pattern does not depend on the number of oscillators in the 2D grid of oscillators. For this, we apply the Ott-Antonsen (OA) approach \cite{ott1,ott2} to study the dynamics of the chimera states from the two-dimensional phase coupled oscillators obtained in Eq. \eqref{eq:6}. Although this OA approach is generally used for the nonidentical systems, it can also be effectively used for homogeneous networks \cite{ott3,ott4}. 
	\par The continuous version of the obtained phase reduced model Eq.\eqref{eq:6} can be written as
	\begin{equation}
	\begin{array}{lcl}
	\frac{\partial\theta(x,y,t)}{\partial t} = \omega'- \lambda \int_{0}^{1}\int_{0}^{1} G(x-x',y-y')\\\\~~~~~~~~~~~~~~ \sin[\theta(x,y,t) - \theta(x',y',t) + \gamma] dx' dy',
	\end{array}
	\end{equation}
	where the coupling kernel $G$ can be written as $G(x-x',y-y') = H[\cos(\sqrt{(x-x')^2 + (y-y')^2})2\pi - \cos(2\pi/N)]$.

	Considering the limit as $N\rightarrow \infty,$ the state of the above system at time $t$ can be described by a probability density function $f(x,y,\theta,t)$, which satisfies the continuity equation 
	\begin{equation}\label{eq: 8}
	\frac{\partial f}{\partial t} + \frac{\partial}{\partial \theta}(fv) = 0,
	\end{equation}
	where 
	\begin{equation}
	v = \frac{d\theta}{dt} = \omega' - \frac{1}{2i}[re^{i\theta} + \bar{r}e^{-i\theta}],
	\end{equation}
	and $r$ is the order parameter given by 
	\begin{equation}
	\begin{array}{lcl}
	r(x,y,t) = \lambda e^{i\gamma} \int_{0}^{1}\int_{0}^{1} G(x-x',y-y')\\\\~~~~~~~~~~~~~~~\int_{0}^{2\pi} e^{-i\theta}f(x',y',\theta,t)d\theta dx' dy'.
	\end{array}	
	\end{equation}
	
 Then the probability density function $f(x,y,\theta,t)$ can be expanded in terms of the Fourier series taking into account the OA ansatz $f_n(x,y,\theta,t)$ = ${h(x,y,t)}^n$ as
	\begin{equation}\label{eq: 11}
	\begin{array}{lcl}
	f(x,y,\theta,t) = \frac{1}{2\pi}\left(1 + \sum_{n=1}^{\infty} {h(x,y,t)}^n e^{in\theta} + \text{c.c.}\right)\\\\~~~~~~~~~~~~~~
	= \frac{1}{2\pi}\left(1 + \sum_{n=1}^{\infty} (h^n e^{in\theta} + \bar{h}^n e^{-in\theta})\right).
	\end{array}
	\end{equation}
	
	Therefore, 	
	\begin{equation}\label{eq: 12}
	\begin{array}{lcl}
	\frac{\partial}{\partial \theta}(fv)=v\frac{\partial f}{\partial \theta} + f\frac{\partial v}{\partial \theta}\\\\~~~~~~~~~
	=\left[\omega' - \frac{1}{2i}(re^{i\theta} + \bar{r}e^{-i\theta})\right]\\\\~~~~~~~~~~~~\left[\frac{1}{2\pi}\sum_{n=1}^{\infty}in\left(h^ne^{in\theta}-\bar{h}^ne^{-in\theta}\right)\right]\\\\~~~~~~~~~+ \left[\frac{1}{2\pi}\left(1+\sum_{n=1}^{\infty}\left(h^ne^{in\theta} + \bar{h}^ne^{-in\theta}\right)\right)\right]\\\\~~~~~~~~~~~~\left[-\frac{1}{2}\left(re^{i\theta} - \bar{r}e^{-i\theta}\right)\right]
	\end{array}
	\end{equation}
	and
	\begin{equation}\label{eq: 13}
	\begin{array}{lcl}
	\frac{\partial f}{\partial t}=\frac{1}{2\pi}\sum_{n=1}^{\infty}\left(nh^{n-1}e^{in\theta}\frac{\partial h}{\partial t} + n\bar{h}^{n-1}e^{-in\theta}\frac{\partial \bar{h}}{\partial t}\right).
	\end{array}	
	\end{equation}
	\\
	Using Eqs. \eqref{eq: 11}--\eqref{eq: 13} from Eq.  \eqref{eq: 8}, we obtain
	\begin{equation*}
		\begin{array}{lcl}
			\frac{1}{2\pi}nh^{n-1}\frac{\partial h}{\partial t}=-\frac{1}{2\pi}[\omega'inh^n - \frac{1}{2i}(ri(n-1)h^{n-1} \\\\~~~~~~~~~ +\bar{r}i(n+1)h^{n+1}) - \frac{1}{2i}\left(irh^{n-1} - i\bar{r}h^{n+1}\right)]
		\end{array}
	\end{equation*}
	\begin{equation}
	\begin{array}{lcl}
	\implies \frac{\partial h}{\partial t}=-i\omega'h + \frac{1}{2}\left(\bar{r}h^2 + r\right),
	\end{array}
	\end{equation}
	
	where
	\begin{equation}
	\begin{array}{lcl}
	r(x,y,t) =\lambda e^{i\gamma} \int_{0}^{1}\int_{0}^{1} G(x-x',y-y')\\\\~~~~~~~~~~~~~~\int_{0}^{2\pi} e^{-i\theta}f(x',y',\theta,t)d\theta dx' dy'\\\\~~~~~~~~~~~
	=\lambda e^{i\gamma} \int_{0}^{1}\int_{0}^{1} G(x-x',y-y')\\\\~~~~~~~\int_{0}^{2\pi} e^{-i\theta}\frac{1}{2\pi}\left(1 + \sum_{n=1}^{\infty}\left(h^ne^{in\theta} + \bar{h}^ne^{-in\theta}\right)\right) d\theta dx' dy'\\\\~~~~~~~~~~~
	=\lambda e^{i\gamma} \int_{0}^{1}\int_{0}^{1} G(x-x',y-y') h(x',y',t) dx'dy'
	\end{array}	
	\end{equation}
	
	We substitute the OA ansatz \cite{chimera_multiplex}, $h = |h|e^{-i\psi}$ in Eq.  \eqref{eq: 11} and get  
	\begin{equation}
	\begin{array}{lcl}
	f(x,y,\theta,t) = \frac{1}{2\pi}\frac{1-|h|^2}{(1-|h|^2)+2|h|[1-\cos(\phi-\psi)]}.
	\end{array}
	\end{equation}
	Here $|h|$ is the maximum value of the phase distribution and $\psi$ is the phase value corresponding to the distribution maximum.
	
	\par
\begin{figure}[ht]
		\centerline{
			\includegraphics[scale=0.4]{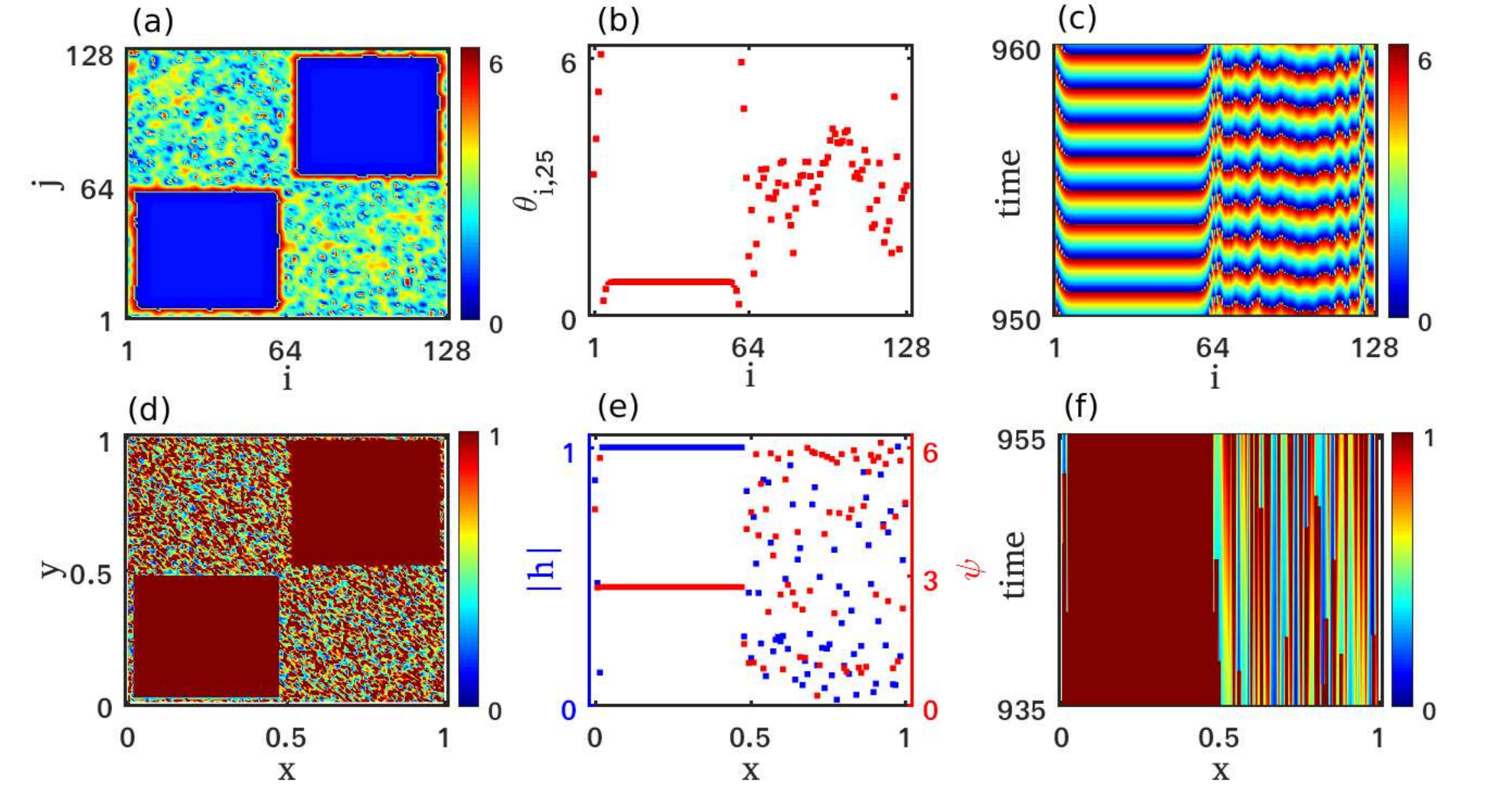}}
		\caption{ Snapshot of  (a) the phase $\theta_{i,j}$ in the 2D lattice, (b) $\theta_{i,25}$ along the horizontal cross-section $j=25$, (c) space-time evolution of the phases $\theta_{i,25}$ along this cross-section line. Snapshot of the (d) maximum $|h(x,y)|$ of the phase distribution in the 2D grid, (e) along the horizontal cross-section line  $|h(x,0.195)|$ (blue dotted) together with the snapshot of $\psi$ (red dotted). (f) Space-time evolution of $|h(x,0.195)|$ reflecting chimera pattern. Here $\alpha=1.0,\beta=-1.5, \tilde a=1.02,\epsilon=0.15,\omega=\alpha-\beta,\gamma=\tan^{-1}\beta $.}
		\label{lsrevisefig2}
	\end{figure}

 Next we move on to analyze the evolution of the phases $\theta_{i,j}, i, j=1, 2, ..., N,$ of the oscillators. Phases $\theta_{i,j}$ of all the oscillators [cf. Eq. \eqref{eq:6}] over the 2D grid and of the $N$ oscillators along the cross-section $j=25$ are plotted in Figs. \ref{lsrevisefig2}(a) and \ref{lsrevisefig2}(b), respectively. These figures clearly validate the existence of chimera pattern obtained in Figs. \ref{lsrevisefig1}(a) and \ref{lsrevisefig1}(b), respectively. Finally, the space-time evolution of the phases $\theta_{i,25}$ is presented in Fig. \ref{lsrevisefig2}(c) that claims stationarity of the chimera pattern.
	\par To confirm the appearance of such chimera states in the 2D grid of locally coupled SL oscillators in the limit of $N \rightarrow \infty$, we have made an attempt to analyze the network behavior through complex OA approach in terms of the absolute value $|h(x,y)|$ corresponding to the maximum of the phase distribution of the oscillators and $\psi$, which is the phase value corresponding to the distribution maximum [cf. Eq. $(16)$]. Figure \ref{lsrevisefig2}(d) represents the snapshot of $|h(x,y)|$ over the 2D grid, whereas snapshot of $|h(x,0.195)|$ (in blue) is shown in the left side panel of Fig. \ref{lsrevisefig2}(e). The emergence of coherent domains are here characterized with $|h(x,y)|= 1$ for which the nearby (positioned) oscillators are phase locked. In contrast, incoherent groups are represented with those oscillators near the sites $x$ and $y$, where $|h(x,y)| < 1$, as the oscillators with those positions have sparsely distributed phases. The snapshot of $\psi$ (in red) is also shown in the right side panel of Fig. \ref{lsrevisefig2}(e). The same phase values for coherence along with the incoherent domain having random distribution in phases corresponding to distribution maximum, is also evident here. Comparing these results with the numerically obtained plots described above, it is quite clear that the analytical treatment [cf. Eqs. $(16)$ and $(17)$] based on the assumption of sufficiently large number of oscillators possessing the same interaction scenario perfectly matches the network behavior (specifically, chimera patterns) realized so far. In addition, space-time plot of $|h(x,0.195)|$ implying stationary evolution of the chimera pattern is portrayed in Fig. \ref{lsrevisefig2}(f).
	
\section{Hindmarsh-Rose  Neuronal Model}

Next we verify the observed chimera states in a more realistic 2D grid of $N\times N$ coupled Hindmarsh-Rose (HR) neuronal oscillators, which are interacting locally through chemical synapses. The mathematical form of the associated coupled network is represented by the following equations:
\begin{equation} \label{eq:hr}
\begin{array}{lcl}
\dot x_{i,j}=ax_{i,j}^2-x_{i,j}^3-y_{i,j}-z_{i,j}+\frac{\epsilon}{4}(v_s-x_{i,j})[\Gamma(x_{i-1,j})\\~~~~~~~~+
\Gamma(x_{i+1,j})+\Gamma(x_{i,j-1})+\Gamma(x_{i,j+1})],\\
\dot y_{i,j}=(a+\alpha)x_{i,j}^2-y_{i,j},\\
\dot z_{i,j}=c(bx_{i,j}-z_{i,j}+e),
\end{array}
\end{equation} for $i,j=1,2,...,N$ with periodic boundary conditions $x_{N+1, j}=x_{1, j}$, $x_{i, N+1}=x_{i, 1}$ together with $x_{0, j}=x_{N, j}$ and $x_{i, 0}=x_{i, N}$. Here, $\epsilon>0$ is the chemical synaptic coupling strength. The variables $x_{i,j}$ represent the membrane potentials of the neuron at the  $(i,j)$-th position of the 2D grid in the coupled HR neuron model whereas the other two variables $y_{i,j}$ and $z_{i,j}$ are associated with the transportation of ions across the membrane through the ion channels. The variables $y_{i,j}$ and $z_{i,j}$ represent the rate of changes of fast current
(associated with Na$^+$ or K$^+$), and  the slow current (associated with Ca$^{2+}$), respectively. This speed is controlled by the modulated value of the parameter $c$. We consider the reversal potential $v_s$ as $v_s=2$ so that $v_s>x_{i,j}(t)$ for all times $t$ and all values $x_{i,j}(t)$, so that the interaction is always excitatory. The chemical synaptic coupling function $\Gamma(x_{i,j})$ is nonlinear and it is described by the sigmoidal input-output function as $\Gamma(x_{i,j})=\frac{1}{1+e^{-\lambda(x_{i,j}-\Theta_s)}}$. The parameter $\lambda=10$ determines the slope of the sigmoidal function and $\Theta_s=-0.25$ is the synaptic firing threshold. We choose the values of the other parameters as $a=2.8,b=9,c=0.001,e=5,\alpha=1.6$, so that in the absence of the synaptic coupling of strength $\epsilon$, the individual neurons exhibit square wave bursting dynamics. \\
\par
\begin{figure}[ht]
\centerline{
\includegraphics[scale=0.5]{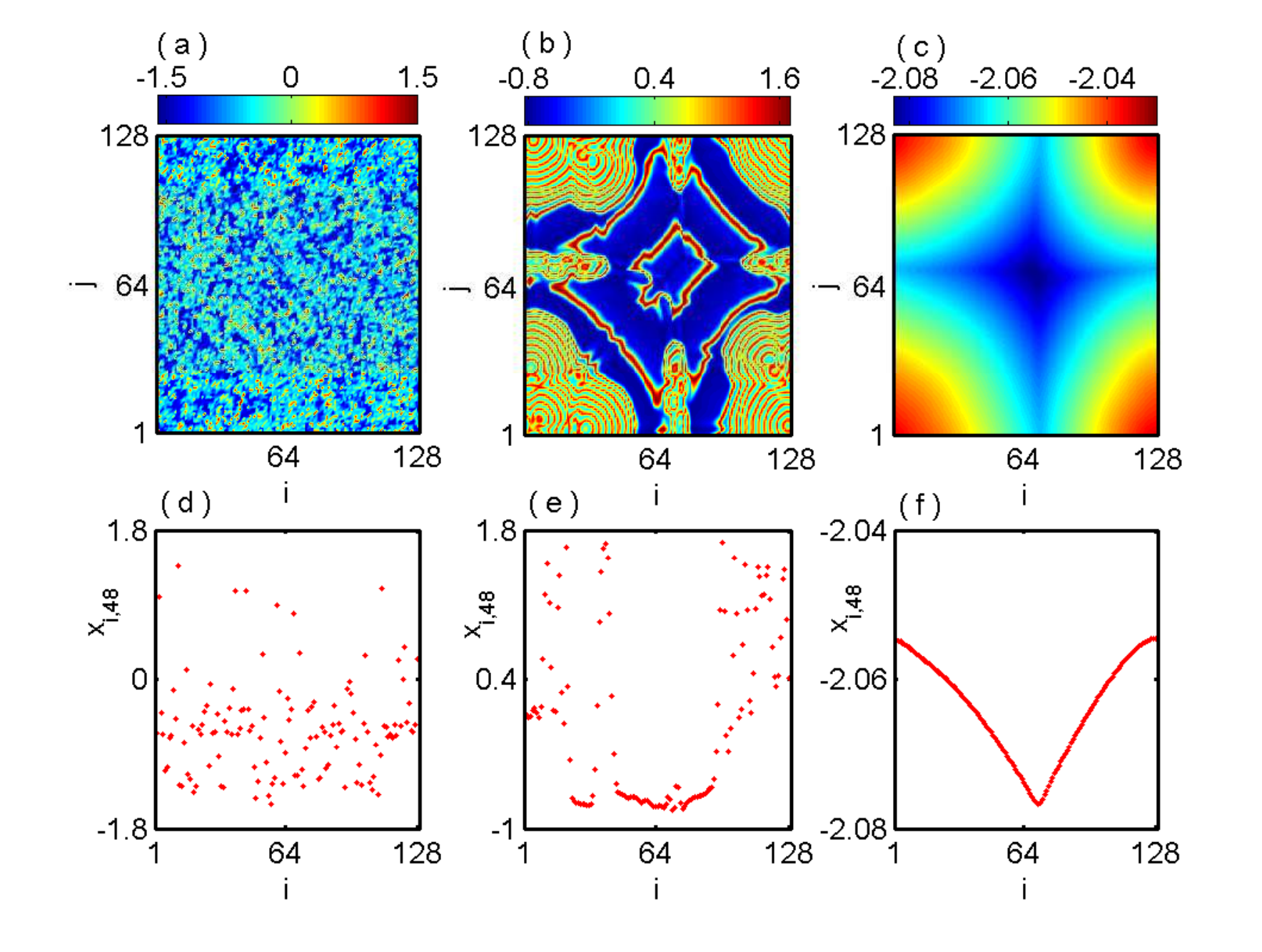}}
\caption{Incoherent, chimera, and coherent states of coupled HR oscillators in a 2D grid. The snapshots of the state variables $x_{i,j}$ in the 2D grid 	at a particular instant $t=1700$ show (a) incoherent state, $\epsilon=0.1$, (b) chimera state,  $\epsilon=1.2$, and (c) coherent state, $\epsilon=2.1$. With horizontal cross-section line  $j=48$ in upper row (a--c), the snapshots in one-dimensional array showing (d) incoherent, (e) chimera, and (f) coherent states. 
}
\label{hrsnap}
\end{figure}

 Now, we numerically study the emergence of several collective dynamical states in the two dimensional grid of coupled network Eq. \eqref{eq:hr} by changing the chemical synaptic coupling strength $\epsilon$. In our simulation, we use the fifth-order Runge-Kutta-Fehlberg algorithm to integrate the above coupled HR systems with a time-step size of $0.01$.  The initial conditions are chosen as $x_{i,j}(0)=0.001[N-(i+j)], y_{i,j}(0)=0.002[N-(i+j)], z_{i,j}(0)=0.003[N-(i+j)]$ for $i,j=1,...,N$ with added small random fluctuations. Figure \ref{hrsnap} shows the several collective dynamical states, which have been identified for different chemical synaptic coupling strengths in the two-dimensionally coupled HR neuron ensemble with $N=128.$ The snapshots of the membrane potentials of all the neurons placed in the 2D grid at a particular time $t=1700$  are plotted in Figs. \ref{hrsnap}(a), \ref{hrsnap}(b), and \ref{hrsnap}(c), representing incoherent, chimera,  and coherent states for coupling strengths $\epsilon=0.1,\epsilon=1.2$, and $\epsilon=2.1$, respectively. Figures \ref{hrsnap}(d)--\ref{hrsnap}(f) depict the corresponding snapshots of neurons in the 2D plane with horizontal cross-section by $j=48$ for incoherent, chimera, and coherent states, respectively. At a lower value of interaction strength $\epsilon=0.1$, all the neurons are randomly distributed resembling a disordered state in the \textit{i-j} space of the 2D grid as shown in Fig. \ref{hrsnap}(a), while the snapshot across a particular value of $j=48$ is given in Fig. \ref{hrsnap}(d). On increasing the coupling strength to $\epsilon=1.2$, the network exhibits chimera pattern as shown in Fig. \ref{hrsnap}(b). With further increment in the value of $\epsilon$ to $\epsilon=2.1$ leads all the oscillators to follow a coherent profile depicted in Fig. \ref{hrsnap}(c).  The color bars in Figs. \ref{hrsnap}(a)--\ref{hrsnap}(c) represent the amplitudes of the membrane potentials $(x_{i,j})$. These three different states appear symmetrically in the \textit{i-j} plane of the two-dimensional grid. 

\par  For HR 2D neuronal network with chemical synaptic interaction, it is rather cumbersome to deal with the Ott-Antonsen approach, though our studies on 2D Stuart-Landau oscillators clearly establish the existence of chimera states in locally coupled nonlinear interactions both analytically and numerically. So further analysis is carried through numerical investigation based on the calculation of order parameter and strength of incoherence. 

\par
\begin{figure}[ht]
\centerline{
\includegraphics[scale=0.54]{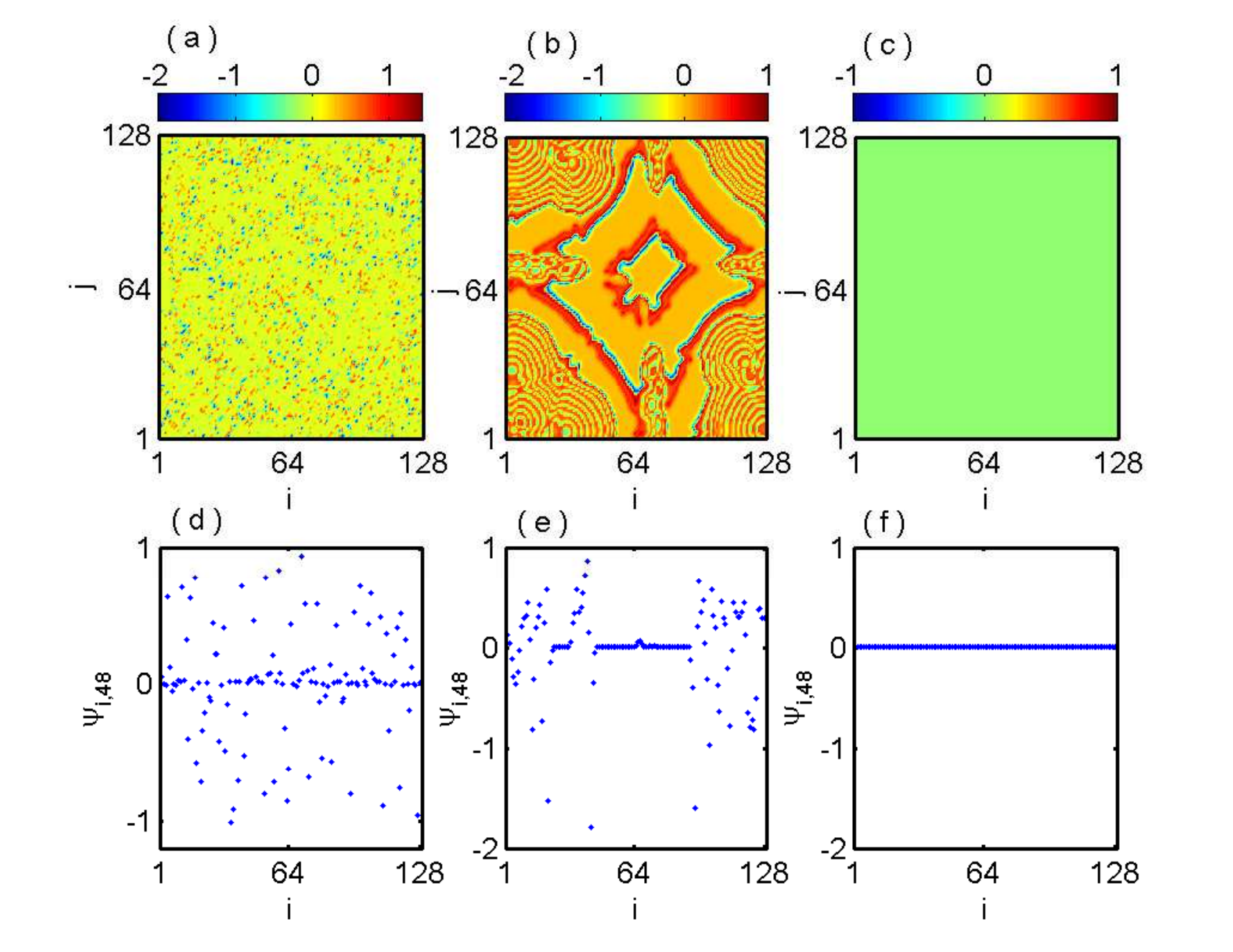}}
\caption{Instantaneous angular frequencies are plotted corresponding to the incoherent, chimera, and coherent states of coupled HR oscillators in the 2D grid. For a particular instant $t=1700,$ the snapshots of the angular frequencies $\Psi_{i,j}$ signifying (a) incoherent, (b) chimera, and (c) coherent states at $\epsilon=0.1, \epsilon=1.2$, and $\epsilon=2.1$, respectively. Along the horizontal cross-section line $j=48$, (d), (e), and (f) represent the snapshots of $\Psi_{i,j}$ in one-dimensional array for incoherent, chimera, and  coherent states, respectively. }
\label{hrfreq}
\end{figure}

 To characterize and distinguish the chimera state from the coherent and incoherent states, we calculate the instantaneous phase and corresponding frequency from the time series of each of the $(i, j)$-th neuron in the 2D grid of coupled HR oscillators. The instantaneous angular frequency \cite{angular_frq} of the $(i, j)$-th neuron is calculated as
\begin{equation}
\begin{array}{lcl}
\Psi_{i, j}= \dot{\phi}_{i, j} =\frac{x_{i, j} \dot{y}_{i, j}-\dot{x}_{i, j} y_{i, j}}{x_{i, j}^2+y_{i, j}^2},
\end{array}
\end{equation}
where $\phi_{i, j}=\mbox{arctan}(y_{i, j}/x_{i, j})$ is the geometric phase for the fast variables $x_{i, j}$ and $y_{i, j}$ of the $(i, j)$-th neuron, which is considered as a good approximation as long as $c$ is small ($<<1$). The angular frequencies corresponding to the neurons in the incoherent domain are randomly scattered, whereas for the coherent domain they remain almost the same.  These angular frequency profiles perfectly distinguish different dynamical behaviors in the network.  Figure \ref{hrfreq} illustrates the instantaneous angular frequencies for incoherent, chimera, and coherent states. This figure confirms the different states in Fig. \ref{hrsnap} by taking the same coupling strength as mentioned therein. The distribution of instantaneous phases is closer to constant line refers the coherent motion, whereas random distribution signifies the incoherent dynamics; on the other hand, for chimera states the distributions are partly random and partly constant.

\begin{figure}[ht]
 \centerline{
 \includegraphics[scale=0.560]{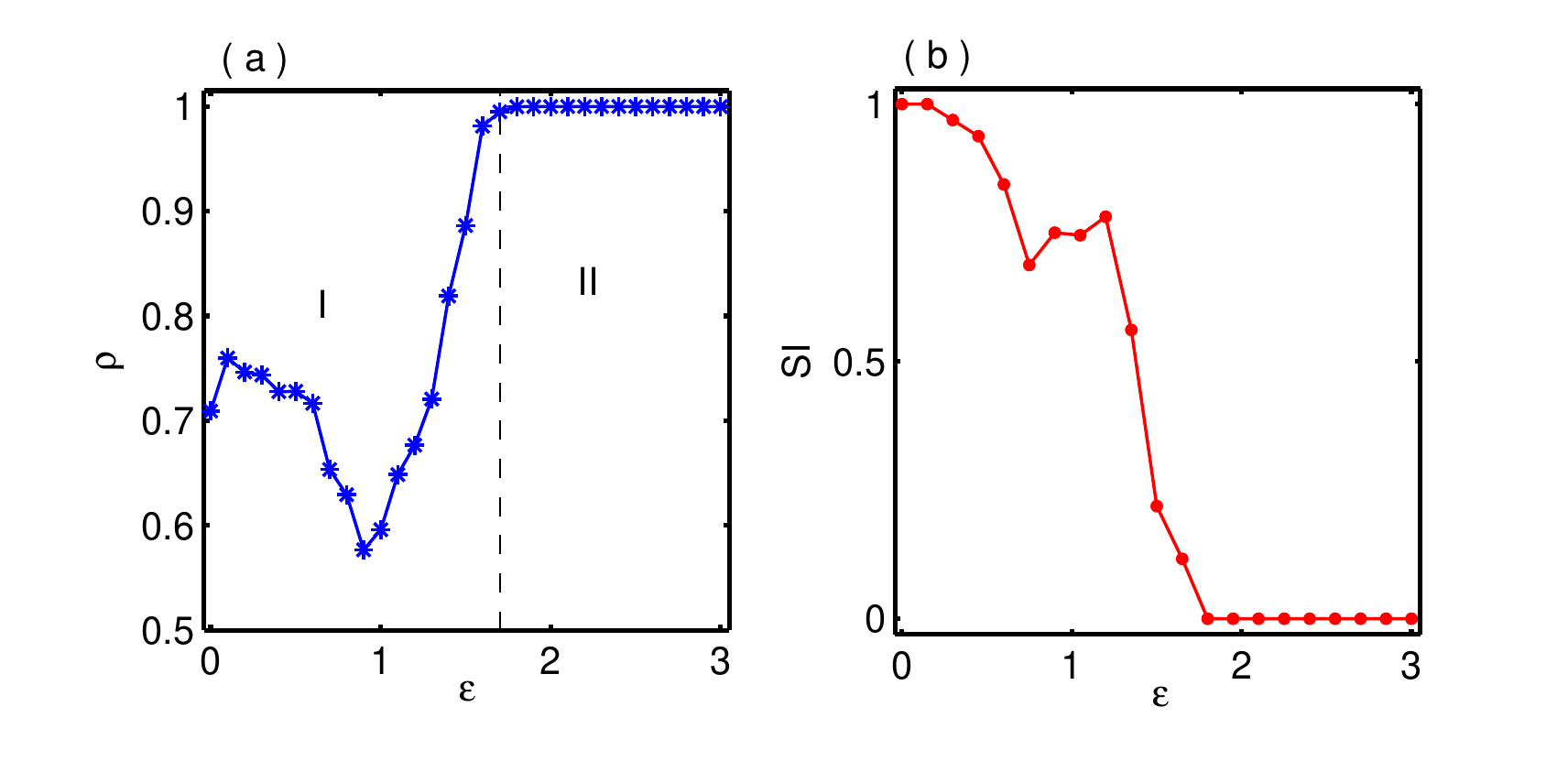}}
 \caption{ (a) Order parameter $\rho$ and (b) strength of incoherence of two-dimensionally coupled HR oscillator are plotted against the chemical synaptic coupling strength $\epsilon$. The regions I and II in (a) indicate the range of incoherent (or chimera) and fully coherent states, respectively. In (b), in the incoherent region the strength of incoherence SI takes a value 1, while in the coherent region it is zero, whereas for chimera states it takes a value between 1 and 0.}
 \label{hropmi}
 \end{figure}

 Next to measure the coherence level of neuronal activity in the two-dimensionally connected neurons, we calculate the Kuramoto order parameter $\rho$, which is defined as the long time average $\rho=\langle \rho(t) \rangle_t$, where $\rho(t)$ is the modulus of complex function 
 \begin{equation}\label{eq:order}
 \begin{array}{lcl}
 z(t)=\rho(t)e^{i \Phi(t)}=\frac{1}{N^2}\sum\limits_{i=1}^{N}\sum\limits_{j=1}^{N} e^{i \phi_{i,j}(t)},
 \end{array}
 \end{equation}
 where $\phi_{i,j}$ is the phase of the $(i,j)$-th neuron and $i=\sqrt{-1}$. The above quantity $\rho$ determines the level of synchronizability with $\rho<<1$ and $\rho\simeq 1$, respectively, characterizing the desynchronized and fully synchronized motion of the coupled network. Figure \ref{hropmi}(a) shows the order parameter $\rho$ with respect to the coupling strength $\epsilon$. The region I corresponds to the zone of incoherent or chimera states, whereas region II stands for the area of fully coherent states.   As order parameter does not distinguish the chimera states from coherent and incoherent states, so to clearly distinguish different collective states further,  we use the statistical measure strength of incoherence (SI) induced by Gopal \textit{et al.} \cite{lakshman_measure} from the time series of the networks. Here we calculate the SI by just taking the horizontal cross section along $j=N_1$ with $i=1,...,N$. First, we introduce the difference variable $w_{i,j}=x_{i,j}-x_{i+1,j}$ and $\langle w \rangle=\frac{1}{N} \sum\limits_{i=1}^{N}w_{i,j}$ for $i=1,...,N$. To distinguish chimera from incoherent and coherent states, we divide the number of oscillators along the horizontal cross section ($j=N_1$) into $p$ (even) bins of equal length $q = \frac{N}{p}$. Then we calculate the local standard deviation which is defined as $\sigma(m)=\left \langle  \sqrt{\frac{1}{q}\sum\limits_{i=q(m-1)+1}^{mq}(w_{i,j}-\langle w \rangle)^2} \right \rangle_t, m=1,...,p$. $\langle...\rangle_t$ represents the long time average and the above quantity $\sigma (m)$ is calculated for every successive $q$ number of oscillators. The strength of incoherence is calculated as 
 \begin{equation}\label{eq:si}	
 \mbox{SI}=1-\frac{\sum\limits_{m=1}^{p}s_m}{p},\;\;\;\;\;\;\;\; s_m=\Theta[\delta-\sigma(m)],
 \end{equation} 
 	where $\Theta(.)$ is the Heaviside step function, and $\delta$ is a predefined threshold which is reasonably small. Here SI $\in [0,1]$ and consequently the values of SI = 1 and SI = 0 characterize the incoherent and coherent states while SI $\in(0,1)$ signifies the chimera states. Variation of SI by changing the chemical synaptic coupling strength $\epsilon$ is shown in Fig. \ref{hropmi}(b) computed along the horizontal cross-section $j=48$. At smaller values of $\epsilon\le 0.15,$ all the neurons exhibit incoherent state where the value of SI is 1. With an increase in the value of $\epsilon$, the chimera state emerges in the region $0.15<\epsilon\le1.8$, where the value of SI lies between 0 and 1. Finally, for $\epsilon> 1.8,$ all the neurons are in coherent state. One can make a similar analysis for any value of $j$ between $1$ and $N$ and verify the dynamical behavior for the 2D grid as well. 

 \section{Rulkov Map}
 We also reveal the above-observed phenomena in yet another neuronal system, namely the Rulkov map \cite{rulkov1,rulkov2}. The mathematical equations of the coupled two-dimensional grid of Rulkov maps are
 \begin{equation}\label{eq:rulkov}
 \begin{array}{lcl} x(n+1)_{i,j}=\frac{\alpha}{1+x(n)_{i,j}^{2}}+y(n)_{i,j}\\+\frac{\epsilon}{4}[v_s-x(n)_{i,j}]\{\Gamma[x(n)_{i-1,j}]
 +\Gamma[x(n)_{i+1,j}]\\+\Gamma[x(n)_{i,j-1}]+\Gamma[x(n)_{i,j+1}]\},\\\\
 y(n+1)_{i,j}=y(n)_{i,j}-\mu[x(n)_{i,j}-\sigma],
\end{array}
   \end{equation} 
  for $i,j=1,2,...,128$ with periodic boundary conditions. The variables $x(n+1)_{i,j}$ represent the membrane potential of the neuron placed at the $(i, j)$-th position of the 2D grid at the discrete time step $n+1$, and $x(n+1)_{i,j}$ is a slow dynamical variable as long as $\mu$ is small ($0<\mu<<1$) and is not explicitly obtained from any biological structure, though  some comparison to gating variables may be drawn. Here,  $\Gamma[x(n)]=\frac{1}{1+e^{-\lambda[x(n)-\Theta_s]}}$ is the chemical synaptic coupling function defined earlier  and $\epsilon>0$ is the chemical synaptic coupling strength. The map displays chaotic behavior for $\alpha>4.0$. The parameter values are fixed at $\alpha=4.1, \mu=0.001, \sigma=-1.6$  for which individual neurons oscillate chaotically and other parameters are fixed at $v_s=2,\Theta_s=-0.25,\lambda=10$.  
 \begin{figure}[ht]
 	\centerline{
 		\includegraphics[scale=0.48]{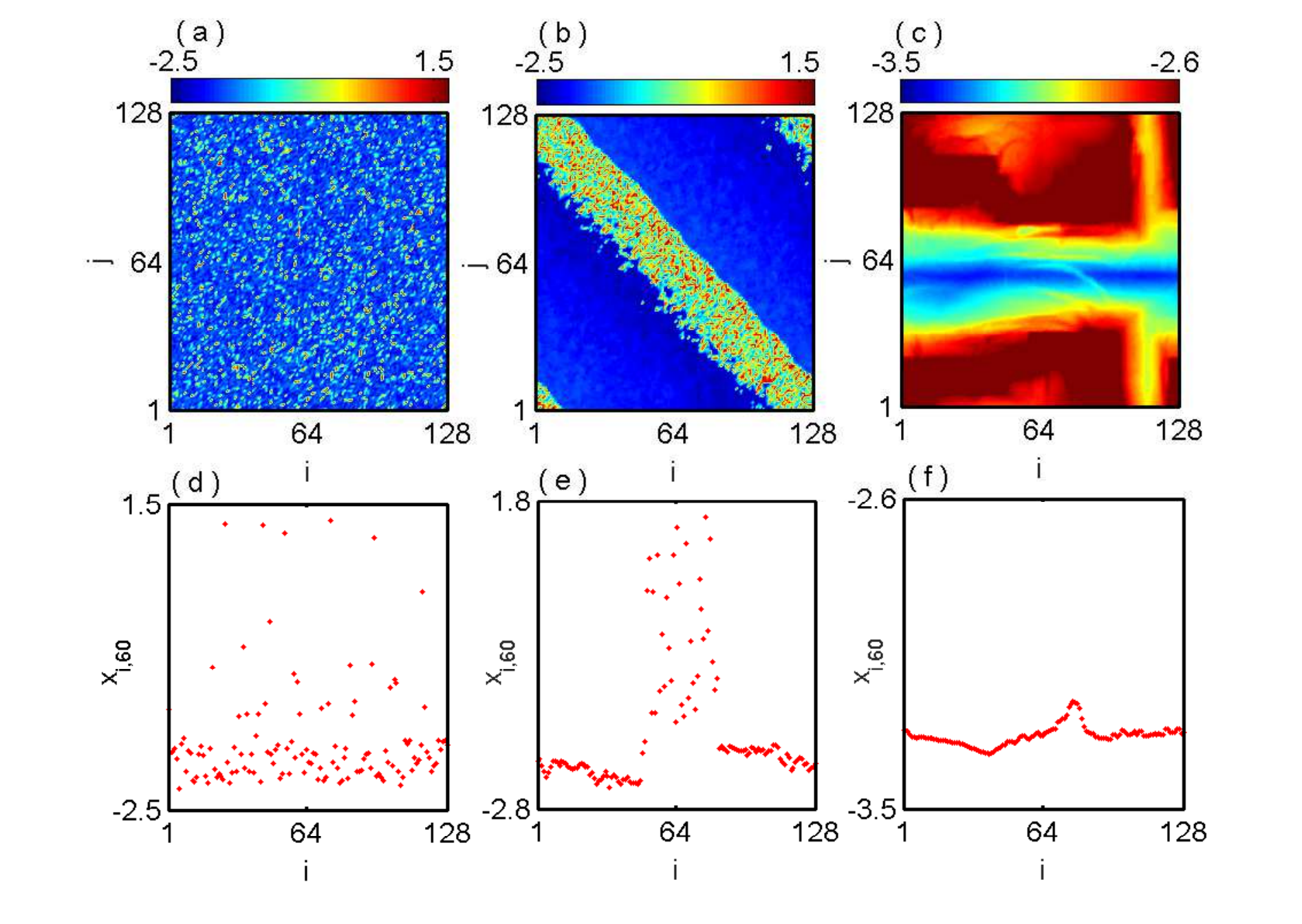}}
 	\caption{Coupled Rulkov maps in a two-dimensional grid: snapshots of membrane potentials showing (a, d) incoherent behavior at $\epsilon=0.004$, (b, e) chimera at $\epsilon=0.2$, and (c, f) coherent behavior at $\epsilon=1.36$ at a particular instant $t=45000$. First row: snapshots of $x(n)$ in the $(i, j)$ plane; second row: snapshots of the states variables $x(n)$ along the horizontal cross-section line $j=60$.}
 	\label{rulksnap}
 \end{figure}
 \par Now we vary the chemical coupling strength $\epsilon$, and explore the different spatiotemporal behaviors of the network Eq. \eqref{eq:rulkov}.  Snapshots at a particular time $(t=45000)$ of the membrane potentials of all the Rulkov maps situated in the 2D grid are shown in Figs. \ref{rulksnap}(a)--\ref{rulksnap}(c) denoting incoherent, chimera  and coherent states for coupling strengths $\epsilon=0.004,\epsilon=0.2$, and $\epsilon=1.36$, respectively. Figures \ref{rulksnap}(d)--\ref{rulksnap}(f) show the snapshots of neurons along the horizontal cross section with $j=60$ of Figs. \ref{rulksnap}(a)--\ref{rulksnap}(c), respectively. At a smaller value of the coupling strength $\epsilon=0.004$, all the membrane potentials are randomly distributed (uncorrelated) which represent an incoherent state as shown in Fig. \ref{rulksnap}(a) with the corresponding snapshot in one-dimensional array along a particular value of $j=60$ is given in Fig. \ref{rulksnap}(d). As the interaction strength is increased to $\epsilon=0.2$, the network admits chimera states, shown in Fig. \ref{rulksnap}(b). For a higher value of  $\epsilon=1.36$, the network of oscillators exhibit a coherent profile as shown in Fig. \ref{rulksnap}(c). 
 
   \begin{figure}[ht]
   \centerline{
   \includegraphics[scale=0.5]{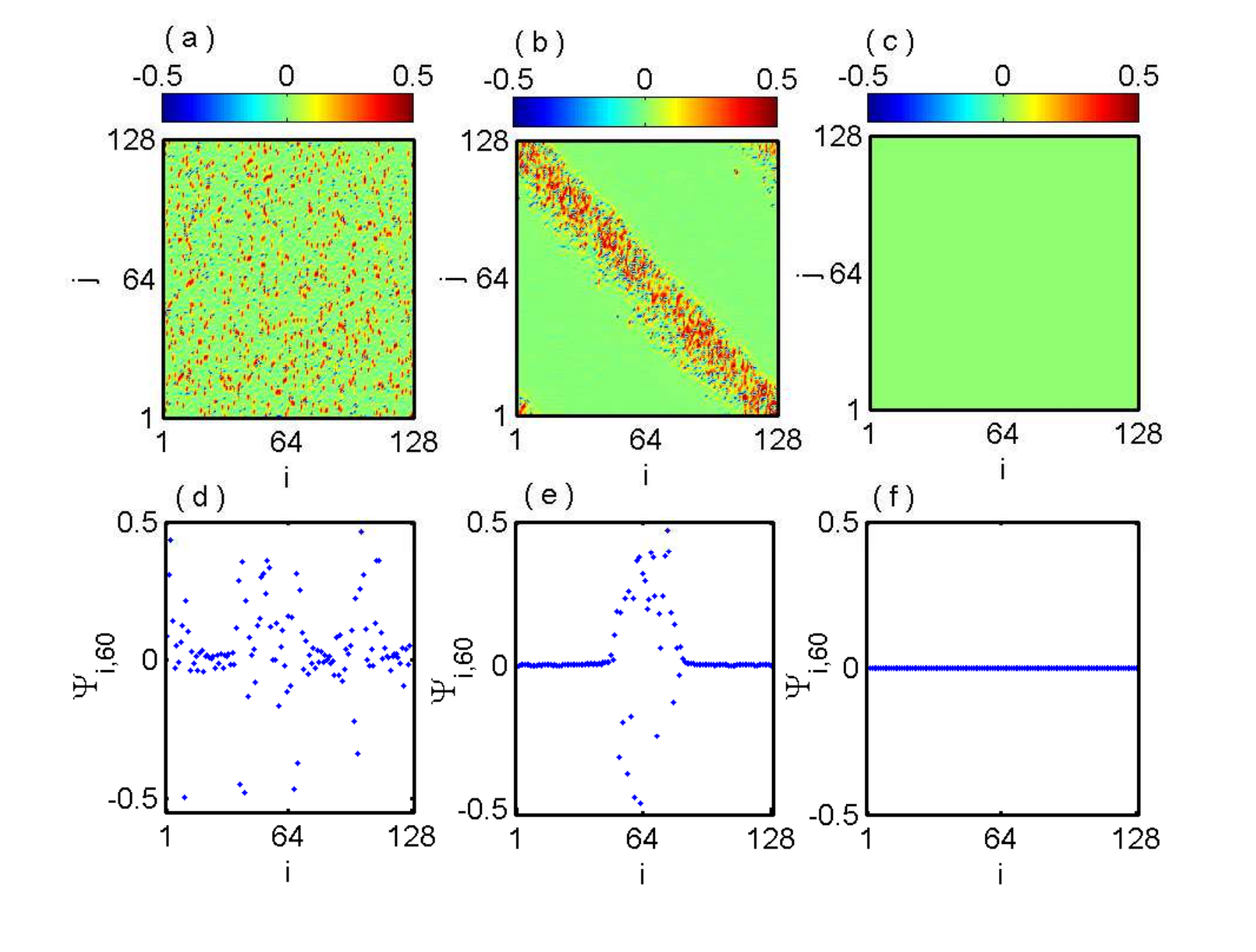}}
   \caption{Snapshots of frequencies $\Psi_{i, j}$ representing incoherent, chimera, and coherent states of the coupled Rulkov maps in a 2D grid are plotted in (a), (b) and (c) for $\epsilon=0.004, \epsilon=0.2$, and $\epsilon=1.36$, respectively,  at $t=45000$. The snapshots of the frequencies $\Psi_{i, j}$ in one-dimensional array along the horizontal cross section $j=60$ are plotted in (d), (e), and (f) for incoherent, chimera, and coherent states, respectively. 
   	 }
   \label{rulkfrq}
   \end{figure}
   
   \begin{figure}[ht]
     \centerline{
     \includegraphics[scale=0.56]{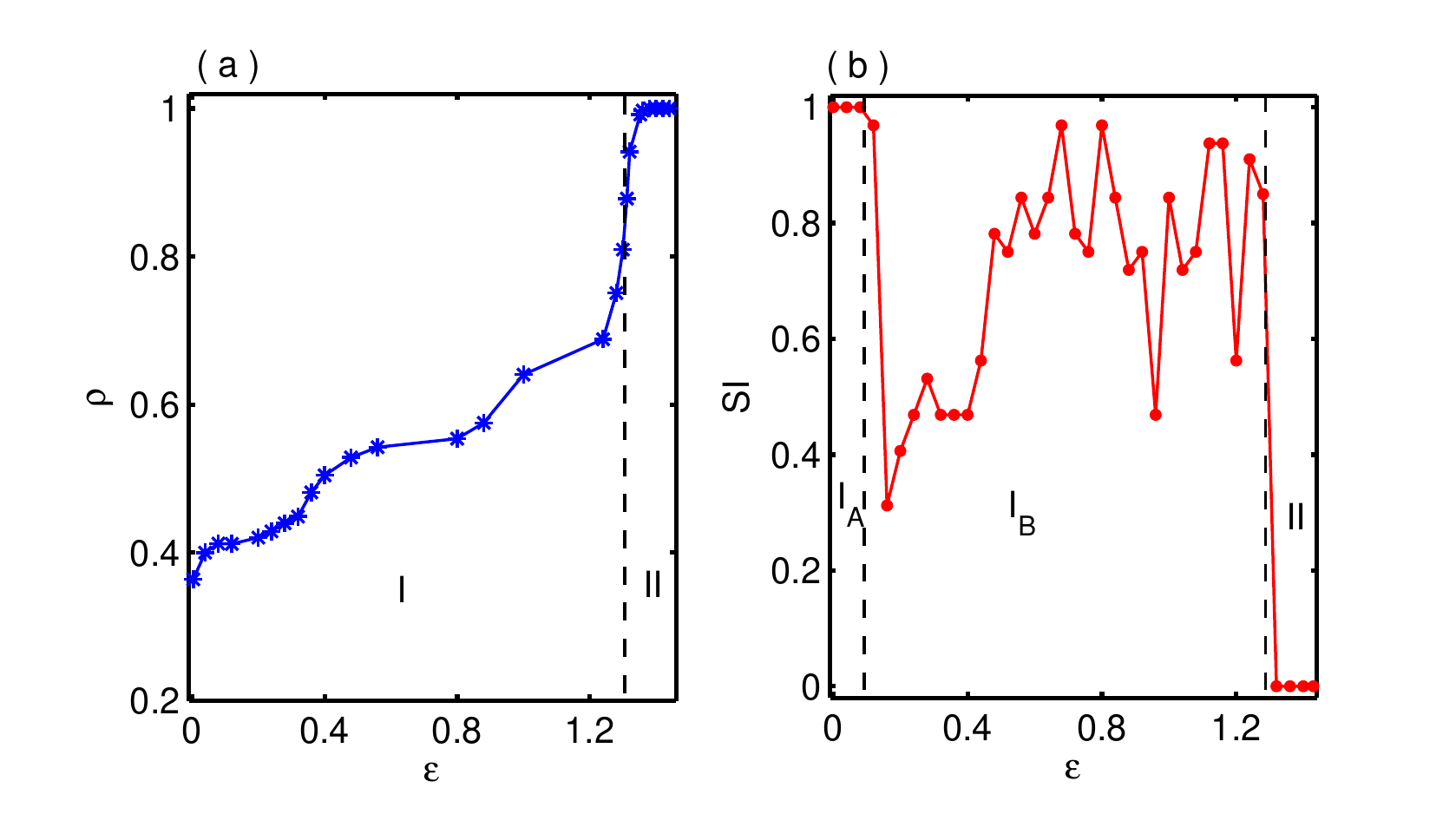}}
     \caption{ Variation of (a) order parameter ($\rho$) and (b) strength of incoherence (SI) by varying the chemical synaptic coupling strength $\epsilon$ in two-dimensionally coupled Rulkov maps.  In (a), regions I and II represents incoherent (or chimera) and coherent states while regions $\mbox I_A$ and $\mbox I_B$ captures the fully desynchronized and chimera states, respectively in (b). } 
     \label{rulkopmi}
     \end{figure}
 
 \par   To compute the oscillator phase and corresponding frequency, we use the analytical signal concept \cite{analytical_signal}, an approach introduced by Gabor \cite{gabor}. An analytical signal $\psi(t)$ is a complex function of time, defined by the amplitude and the phase of an arbitrary variable $s(t)$ as 
  \begin{equation}
  \psi(t)=s(t)+i\tilde{s}(t)=R(t)e^{i\phi(t)}, \;\;\;\;\;\;\; i=\sqrt{-1},
   \end{equation}
 where the function $\tilde{s}(t)$ is the Hilbert transform of $s(t)$ given by
  \begin{equation}
  \tilde{s}(t)=\pi^{-1}\text{P.V.}\int_{-\infty}^{\infty}\frac{s(\tau)}{t-\tau}d\tau,
   \end{equation}
   P.V. means that the integral is taken in the sense of the Cauchy principal value. The instantaneous amplitude $R_{i, j}(t)$ and the instantaneous phase $\phi_{i, j}(t)$ of the variable $s_{i, j}(t)$ of the ${(i, j)}$-th oscillator can be uniquely defined as    $R_{i, j}(t)=\sqrt{s_{i, j}(t)^2+\tilde{s}_{i, j}(t)^2}$,  $\phi_{i, j}=\tan^{-1}\frac{\tilde{s}_{i, j}(t)}{s_{i, j}(t)}$. The corresponding frequency is computed as $\Psi_{i, j}=\dot \phi_{i, j}$, where the dot denotes  derivative with respect to time.
    Figure \ref{rulkfrq} shows the instantaneous angular frequencies corresponding to the incoherent, chimera and coherent states shown in Fig. \ref{rulksnap}. 
\par Next we calculate  the order parameter $\rho$ [using Eq. \eqref{eq:order}] and strength of incoherence  [from Eq. \eqref{eq:si}] to quantify the degree of synchronization and distinguish different  states in the two-dimensional grid of coupled oscillators, respectively. As mentioned earlier, SI takes the value close to $0$ and order parameter $\rho$ takes the unit value for coherent states.  
 
In Figs. \ref{rulkopmi}(a) and \ref{rulkopmi}(b), the variation of the order parameter $\rho$ and SI are shown, respectively, by varying the chemical synaptic coupling strength $\epsilon$.  Here SI is computed by taking horizontal cross-section through $j=60$ in 2D grid network.   The separated regions I and II are marked as in Fig. \ref{rulkopmi}(a) for the existence of incoherent (or chimera) and fully coherent states, respectively. The lower values (close to $0$) of order parameter $\rho$ implies the completely desynchronized dynamics and for certain increased values of the synaptic coupling strength, $\rho\in(0,1)$, which is the region of incoherent states marked as I=$\{\epsilon:0\le\epsilon<1.32 \}$, while the higher value of $\rho \approx 1$ signifies the perfect synchrony marked as II=$\{\epsilon:\epsilon\ge 1.32 \}$. The chimera state can be distinguished from incoherent and coherent state by calculating the value of SI.  At the lower values of the synaptic coupling strength $\epsilon$, SI takes the value $1$ which signifies that all the neurons are fully incoherent and are sustained up to the certain threshold of $\epsilon$ marked as I$_A=\{\epsilon:0\le\epsilon<0.12 \}$ in Fig.~\ref{rulkopmi}(b). With an increase of synaptic coupling strength $\epsilon$ beyond a critical value SI lies in $(0,1)$, which characterizes the chimera states marked as I$_B=\{\epsilon:0.12\le\epsilon<1.32 \}$.  Further increasing the coupling strength $(\epsilon\ge1.32)$, SI takes the values $0$, which indicates all the neurons are in completely coherent states and persists up to the higher values of $\epsilon$ shown in Fig.~\ref{rulkopmi}(b).

\section{Conclusion}
Coexistence of coherence (synchronization) and incoherence (desynchronization) in coupled identical oscillators (popularly termed as chimera) is very much related to neuronal network systems. For instance, in various types of brain diseases such as Parkinson's disease, epileptic seizures, schizophrenia, etc., this exceptional state has ample relevance. Again, this state is also  connected to diverse neuronal developments, such as the unihemispheric slow-wave sleep of some aquatic mammals. Two-dimensional lattice as the interactional platform for neurons in the brain is really evident and the emergence of diverse collective behaviors in coupled neuronal systems under this framework is still unavailable.   
\par In this paper, we have studied the existence of chimera states in two-dimensional coupled systems by considering nonlinear coupling functions to cast the interaction between the nearest-neighbor dynamical units. Taking ``pull-push'' type of nonlinear coupling form in the network of Stuart - Landau oscillators, we observed chimera patterns and analytically verified the obtained results through Ott-Antonsen approach. The analytical results very well match with the obtained numerical results. Here it has been shown that the presence of nonlinearity in the coupling form plays a crucial role for the emergence of chimera states in 2D lattices of locally coupled oscillators which remove the restriction of nonlocality in the coupling topology. Different types of chimera states may exist for different choices of nonlinear coupling functions in the 2D lattice of locally coupled oscillators. By taking realistic communicating medium among neurons, namely chemical synaptic function, we investigated the chimera states in 2D network of neural oscillators. We provided  evidence for the existence of such fascinating complex patterns in two paradigmatic neuronal systems, one continuous time dynamical system and the other one is a discrete time system. For the former case, we consider the Hindmarsh-Rose neural oscillator and for the latter we illustrate  with the Rulkov map. We confirm the appearance of chimera states in these systems as a link between incoherence and coherence by plotting instantaneous angular frequency and Kuramoto order parameters and strength of incoherence measure is used to characterize the incoherent, chimera, and coherent states. Our present study is expected to provide a better understanding of several neuronal developments in which synchronization and desynchronization coexist.

\noindent \textbf{Acknowledgments} \\
D.G. was supported by SERB-DST (Department of Science and Technology), Government of India (Project No. EMR/2016/001039). M.L. is also  supported by a SERB-DST research project (Project No. EMR/2014/001076) and a NASI Platinum Jubilee Senior Scientist Fellowship. D.G. thanks Nikita S. Frolov for helpful discussions.
\\  

  	\section*{Appendix: Transition scenario from incoherence to coherence in the presence of linear coupling function}
  	In this appendix we will show that the presence of linear coupling function instead of nonlinear function in the 2D grid of locally coupled networks never produces chimera states, rather it leads to coherent or incoherent states. To verify this, we consider the three systems, namely SL oscillators, HR system and Rulkov model which are coupled locally in 2D network with linear coupling functions.   
  	
  	\subsection{Coupled SL network}
  	We replace the nonlinear coupling function by linear coupling function in Eq. (\ref{eq:1}) and the corresponding governing equations for the 2D coupled SL network becomes  	
  	\begin{equation} \label{eq:LSlinear}
  	\begin{array}{lcl}
  	\dot{z}_{i,j} = (1+i\alpha)z_{i,j} - (1+i\beta)|z_{i,j}|^2z_{i,j} + \frac{\epsilon}{4}[z_{i-1,j}\\\\\hspace{25pt}+z_{i+1,j} + z_{i,j-1} + z_{i,j+1} - 4z_{i,j}],
  	\end{array}
  	\end{equation}
  	with subscripts $i,j=1,2,...,N$ obeying periodic boundary conditions $z_{N+1,j}=z_{1,j}, z_{i, N+1}=z_{i,1}$ and $z_{0,j}=z_{N,j}, z_{i,0}=z_{i,N}$. Here  $\epsilon$ is the coupling constant.
  	
  	Using similar approach as in Sec. IIIA, the phase reduced model with linear coupling function becomes  
  	\begin{equation}\label{eq:app1}
  	\dot{\theta}_{i,j} = \omega''_{i,j} - \lambda' \sum_{m=1}^{N}\sum_{n=1}^{N} A_{ijmn} \sin(\theta_{i,j} - \theta_{m,n} + \gamma),
  	\end{equation}
  	whose continuous version in the limit $N \rightarrow \infty$ becomes
  	\begin{equation}\label{eq:app2}
  	\begin{array}{lcl}
  	\frac{\partial\theta(x,y,t)}{\partial t} = \omega''- \lambda' \int_{0}^{1}\int_{0}^{1} G(x-x',y-y')\\\\~~~~~~~~~~~~~~ \sin(\theta(x,y,t) - \theta(x',y',t) + \gamma) dx' dy',
  	\end{array}
  	\end{equation}
  	where $\omega''_{i,j}$ = $\omega_{i,j} + \epsilon\beta$, $\lambda'$ = $\frac{\epsilon}{4}\sqrt{1+\beta^2}$ and all other expressions and parameter values are same as in Sec. III. Following similar approach, the equation for OA ansatz $h$ can be written as 
  	\begin{equation}
  	\begin{array}{lcl}\label{appen1}
  	\frac{\partial h}{\partial t}=-i\omega''h + \frac{1}{2}\left(\bar{r}h^2 + r\right),
  	\end{array}
  	\end{equation}
  	where 	 
  	\begin{equation}
  	\begin{array}{lcl}
  	r(x,y,t) =\lambda' e^{i\gamma} \int_{0}^{1}\int_{0}^{1} G(x-x',y-y') h(x',y',t) dx'dy'.
  	\end{array}	
  	\end{equation}
  	
  	\begin{figure}[ht]
  		\centerline{
  			\includegraphics[scale=0.56]{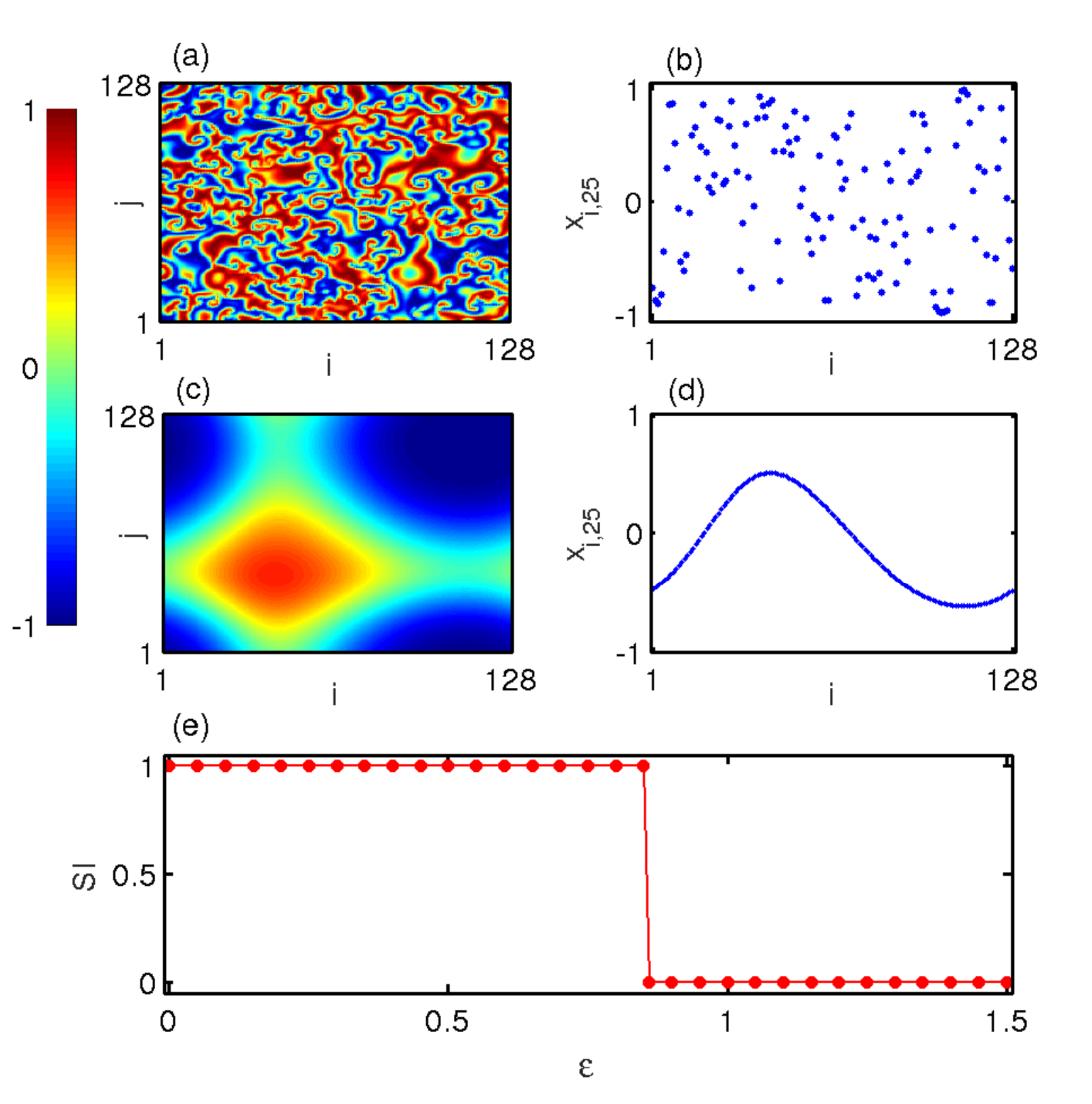}}
  		\caption{Two dimensional grid of SL network in the presence of linear coupling function. Snapshots of $x_{i,j}$ for incoherent states (a) for $\epsilon=0.85$ and coherent states (c) for $\epsilon=0.9$. (b) and (d) represent the snapshots along the horizontal cross-section $j=25$ in $i-j$ plane corresponding to the figures (a) and (c). The variation of SI against the coupling strength $\epsilon$ is shown in (e).} 
  		\label{LS_appen_1}
  	\end{figure}
  	\par The transition scenario from incoherent to coherent states in two dimensional grid of locally coupled network with linear scalar diffusive interaction functions are investigated numerically and analytically. For the continuous variation of the coupling parameter $\epsilon$, we observed the direct transition from incoherent to coherent state in 2D coupled network. The snapshots of the state variables $x_{i,j}$ in the 2D coupled network at a particular instant are plotted in Figs. \ref{LS_appen_1}(a) and \ref{LS_appen_1}(c). For a lower coupling strength $\epsilon=0.85$, all the oscillators in the coupled networks are randomly distributed in the \textit{i-j} plane which indicates the incoherent state of the 2D grid as shown in Fig.~\ref{LS_appen_1}(a) and the corresponding color bar represents the variation of the $x_{i,j}$. The snapshot along the horizontal cross section $j=25$ is plotted in Fig.~\ref{LS_appen_1}(b). The snapshot of coherent state for a certain increased coupling strength $\epsilon=0.9$ is depicted in Fig. \ref{LS_appen_1}(c) and the snapshot through the horizontal cross section $j=25$ in Fig.~\ref{LS_appen_1}(d) shows the smooth profile of the dynamical units which represents a coherent state in the 2D network. To confirm the direct transition scenario from incoherent to coherent states, we plot the strength of incoherence (SI) (discussed in the main text) along the horizontal cross-section $j=25$ in Fig.~\ref{LS_appen_1}(e) with respect to the coupling strength $\epsilon$. From the variation of SI, it is clear that SI takes the value ``1" for the incoherent state upto a certain threshold value of $\epsilon=0.85$, after which for the next increment of $\epsilon$, SI converge to``0" value at $\epsilon=0.86$ and beyond which signifies the coherent states. 
  	
  	\begin{figure}[ht]
  		\centerline{
  			\includegraphics[scale=0.5]{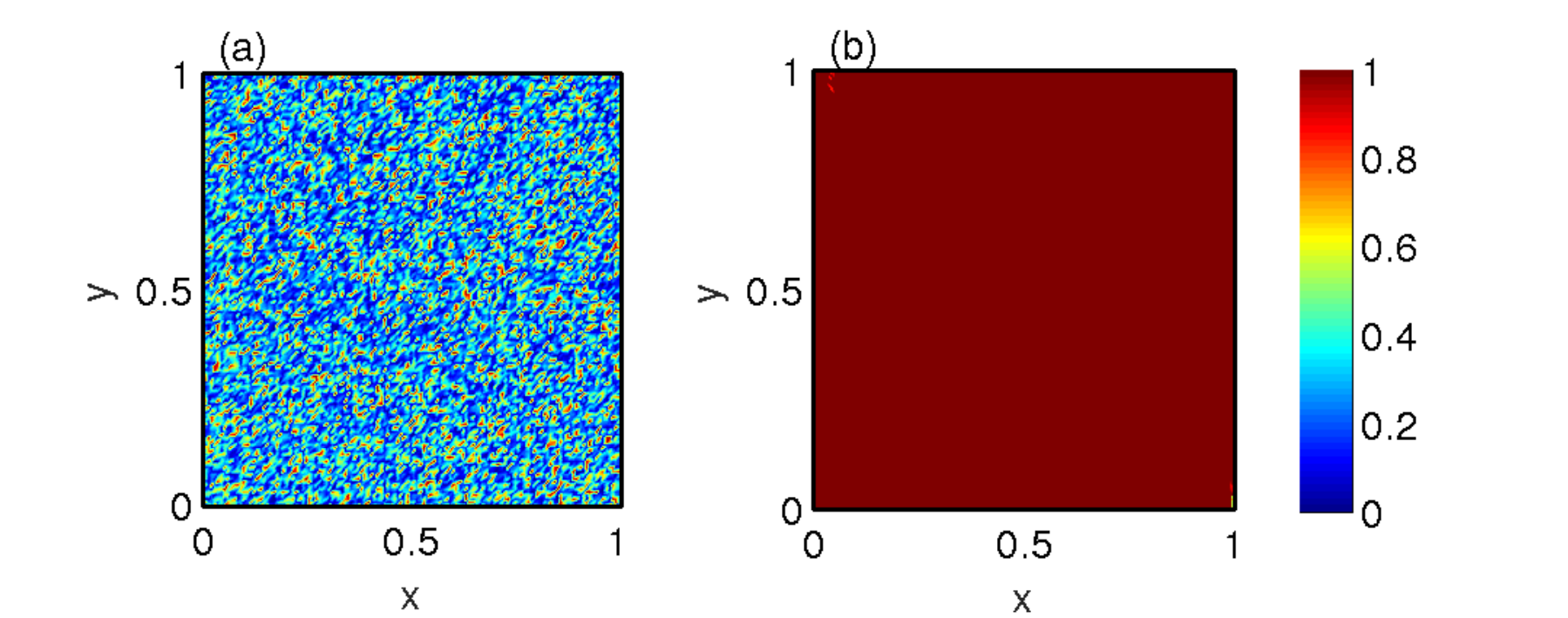}}
  		\caption{(a) and (b) represents the snapshot of absolute values of $h(x,y)$ for incoherent and coherent states corresponding to the Fig. \ref{LS_appen_1}(a) and \ref{LS_appen_1}(c) at the coupling values $\epsilon=0.85$ and $\epsilon=0.9$ respectively.} 
  		\label{LS_appen_2}
  	\end{figure}
  	\par We have also analytically verified the above type of transitions by obtaining the quantity $h(x,y)$ through the complex OA approach in the thermodynamical limit as $N\rightarrow\infty$. In the presence of linear scalar diffusive interaction function of 2D locally coupled LS network, the obtained functional values of $h(x,y)$ is presented in Eq.(\ref{appen1}). The absolute values of $h(x,y)$ for incoherent states and coherent states are plotted in Figs. \ref{LS_appen_2}(a) and \ref{LS_appen_2}(b) in the \textit{x-y} plane corresponding to the coupling values as in Figs. \ref{LS_appen_1}(a) and \ref{LS_appen_1}(c).
  	\subsection{Coupled HR network}
\begin{figure}[ht]
  		\centerline{
  			\includegraphics[scale=0.54]{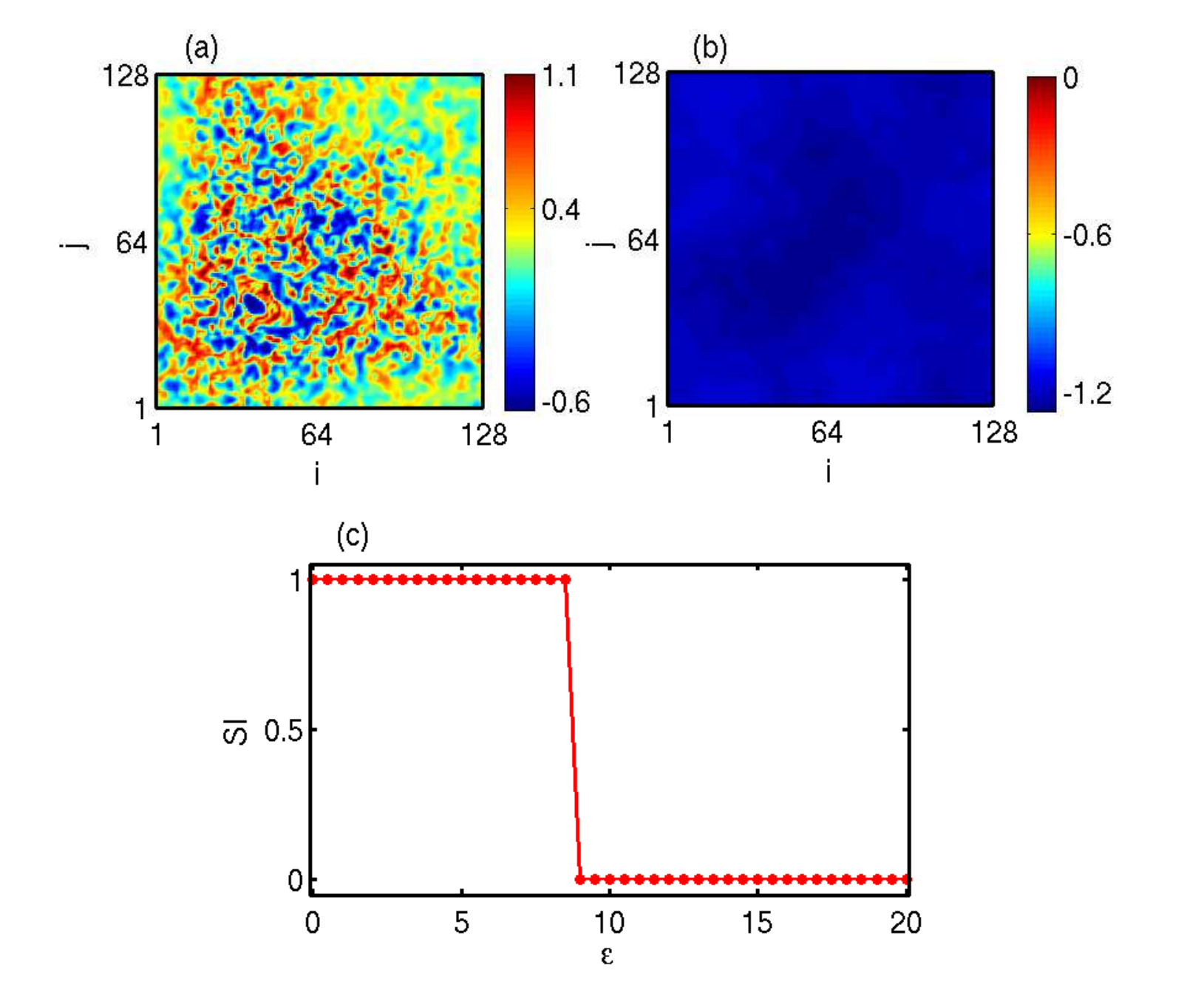}}
  		\caption{Two dimensional locally coupled HR network with electrical synapses in the presence of linear coupling function. (a) Incoherent state, for  $\epsilon=3$, (b) coherent state, for $\epsilon=15$ and the variation of SI characterization is shown in (c) with respect to $\epsilon$.} 
  		\label{HR_appen_1}
  	\end{figure}

  	Next, we verify the results in 2D grid of locally coupled HR network with electric synapses (scalar diffusive coupling function). The dynamical equation of the above network Eq. \eqref{eq:hr} becomes
  	\begin{equation} \label{eq:hr_linear}
  	\begin{array}{lcl}
  	\dot x_{i,j}=ax_{i,j}^2-x_{i,j}^3-y_{i,j}-z_{i,j}+\frac{\epsilon}{4}[x_{i-1,j}+x_{i+1,j}+\\~~~~~~~~x_{i,j-1}+x_{i,j+1}-4x_{i,j}],\\
  	\dot y_{i,j}=(a+\alpha)x_{i,j}^2-y_{i,j},\\
  	\dot z_{i,j}=c(bx_{i,j}-z_{i,j}+e),
  	\end{array}
  	\end{equation} for $i,j=1,2,...,N$. Here $\epsilon>0$ denotes the electrical synaptic coupling strength. The snapshots of the amplitudes of each of the neurons in the 2D locally coupled HR network at a particular time are illustrated in Figs. \ref{HR_appen_1}(a) and \ref{HR_appen_1}(b) for incoherent and coherent states at $\epsilon=3.0$ and $\epsilon=15.0$, respectively. To calculate SI, the cross-section along $j=48$ is taken and the variation of SI for different values of electrical coupling strength $\epsilon$ is shown in Fig. \ref{HR_appen_1}(c). From this figure, it is observed that upto a certain value of the coupling strength $\epsilon=8.5$, SI takes value ``1", which signifies the existence of incoherent states and for further little increment of $\epsilon$, SI gives ``0" value at $\epsilon=9.0$ which corresponds to the appearance of the coherent state.
  	
  	\subsection{Coupled Rulkov Map}
\begin{figure}[ht]
  		\centerline{
  			\includegraphics[scale=0.5]{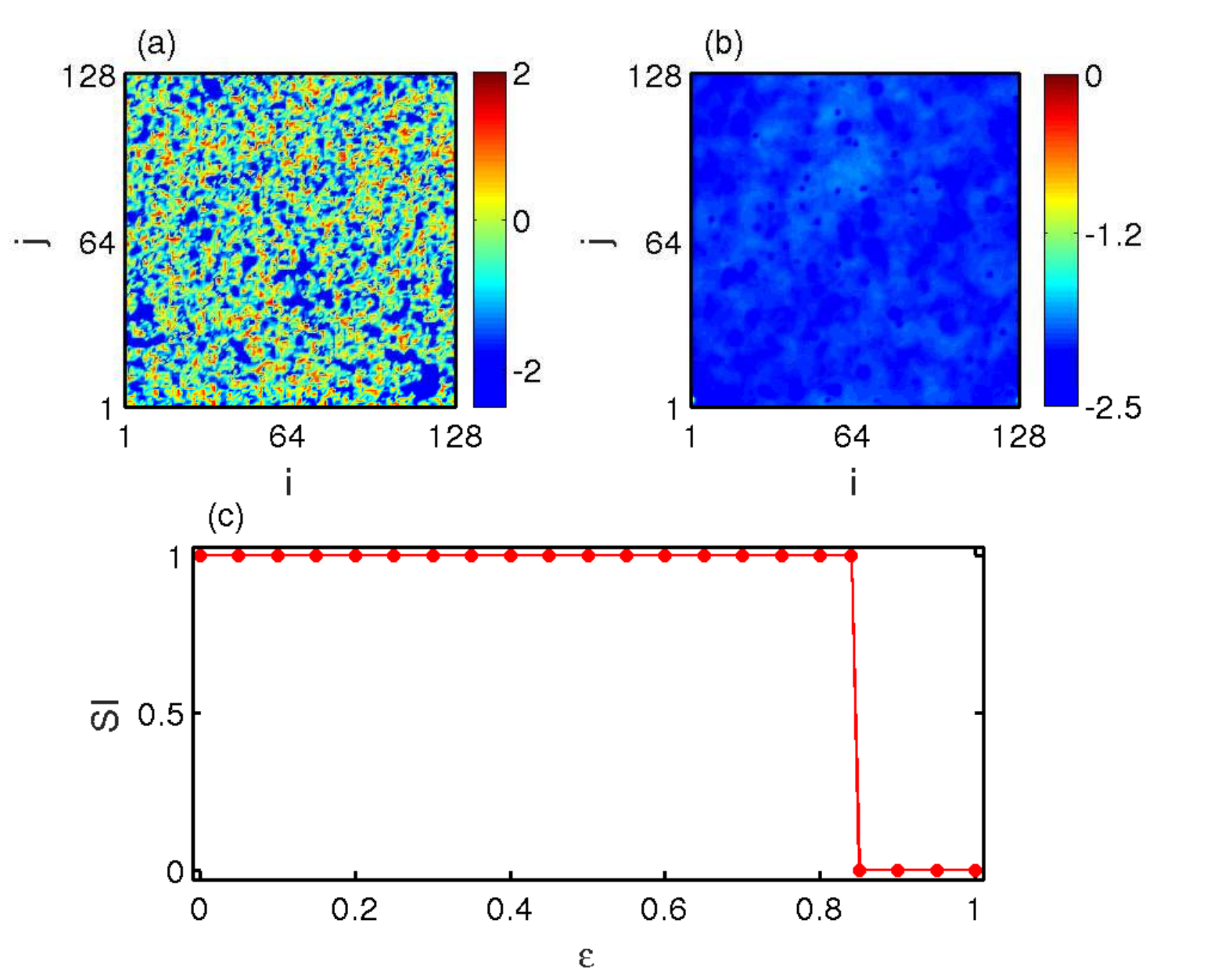}}
  		\caption{ Two dimensional locally coupled Rulkov network with electrical synapses. (a) Incoherent state at  $\epsilon=0.7$, (b) coharent state at $\epsilon=0.9$, and (c) variation of SI with respect to $\epsilon$.}
  		\label{Rulkov_appen_1}
  	\end{figure}

  	The set of mathematical equations of locally coupled Rulkov neurons with electrical synapses in 2D grid network is described by  		 
  	\begin{equation}\label{eq:rulkov_linear}
  	\begin{array}{lcl} x(n+1)_{i,j}=\frac{\alpha}{1+x(n)_{i,j}^{2}}+y(n)_{i,j}+\frac{\epsilon}{4}[x(n)_{i-1,j}
  	+x(n)_{i+1,j}\\~~~~~~~~~~~~~~~~+x(n)_{i,j-1}+x(n)_{i,j+1}-4x(n)_{i,j}],\\\\
  	y(n+1)_{i,j}=y(n)_{i,j}-\mu[x(n)_{i,j}-\sigma],
  	\end{array}
  	\end{equation}  	 
  	for $i,j=1,2,...,N$ and and $\epsilon$ is the coupling strength and all other parameters carry the same meanings as in Sec. V. Figures \ref{Rulkov_appen_1} (a) and \ref{Rulkov_appen_1}(b) represent the incoherent and coherent dynamics for two synaptic coupling strengths at $\epsilon=0.7$ and $\epsilon=0.9$, respectively. The smooth transition from incoherent to coherent states is characterized by the variation of the SI measurement in Fig. \ref{Rulkov_appen_1}(c) with respect to the coupling strength $\epsilon$. Here SI is calculated by taking the cross section along $j=60$ from the \textit{i-j} plane.
  


\begin{thebibliography}{25}

	\bibitem{uhsws1} N. C. Rattenborg, C. J. Amlaner, and S. L. Lima, \textit{Neurosci. Biobehav. Rev.} {\bf 24}, 817 (2000).
	\bibitem{uhsws2} N. C. Rattenborg, \textit{Naturwissenschaften} {\bf 93}, 413 (2006).
     \bibitem{kuramoto} Y. Kuramoto, and D. Battogtokh, \textit{Nonlinear  Phenom. Complex Syst.} {\bf 5}, 380 (2002).
	\bibitem{strogatz} D. M. Abrams, and S. H. Strogatz, \textit{Phys. Rev. Lett.} {\bf 93}, 174102 (2004).	
	\bibitem{brain_disease1} R. Levy, W. D. Hutchison, A. M. Lozano, and J. O. Dostrovsky, \textit{J. Neurosci.} {\bf 20}, 7766 (2000).
	\bibitem{brain_disease2} G. F. Ayala, M. Dichter, R. J. Gumnit, H. Matsumoto, and W. A. Spencer, \textit{Brain Res.} {\bf 52}, 1 (1973).
	\bibitem{chimera_rev} M. J. Panaggio, and D. M. Abrams, \textit{Nonlinearity} {\bf 28}, R67 (2015).
	\bibitem{epl_rev} B. K. Bera, S. Majhi, D. Ghosh and M. Perc, \textit{Europhys. Letts.} {\bf 118}, 10001 (2017).	
	\bibitem{limit} S. Ulonska, I. Omelchenko, A. Zakharova, and E. Sch\"{o}ll, \textit{Chaos} {\bf 26}, 094825 (2016).	
	\bibitem{limit2} B. K. Bera, D. Ghosh, P. Parmananda, G. V. Osipov,	and S. K. Dana, \textit{Chaos} {\bf 27}, 073108 (2017).		
	\bibitem{chaotic} C. Gu, G. St-Yves, and J. Davidsen, \textit{Phys. Rev. Lett.} {\bf 111}, 134101 (2013).
	\bibitem{chaotic_map} I. Omelchenko, Y. Maistrenko, P., H\"{o}vel, and E. Sch\"{o}ll, \textit{Phys. Rev. Lett.} {\bf 106}, 234102 (2011).
	\bibitem{lakshman_measure} R. Gopal, V. K. Chandrasekar, A. Venkatesan, and M. Lakshmanan, \textit{Phys. Rev. E} {\bf 89}, 052914 (2014).
	\bibitem{bs_chimera} S. Rakshit, B. K. Bera, M. Perc and D. Ghosh, \textit{Sci. Rep.} {\bf 7}, 2412 (2017).	
	\bibitem{hr_bera1} B. K. Bera, D. Ghosh, and M. Lakshmanan, \textit{Phys. Rev. E} {\bf 93}, 012205 (2016).
	\bibitem{chimera_modular} J. Hizanidis, N. E. Kouvaris, G. Zamora-L\'{o}pez, A. D\'{i}az-Guilera, and C. G. Antonopoulos, \textit{Sci. Rep.} {\bf 6}, 19845 (2016).
	\bibitem{hr_ijbc} J. Hizanidis, V. Kanas, A. Bezerianos, and T. Bountis, \textit{Int. J. Bifurcat. Chaos} {\bf 24}, 1450030 (2014).
	\bibitem{global1} A. Yeldesbay, A. Pikovsky, and M. Rosenblum, \textit{Phys. Rev. Lett.} {\bf 112}, 144103 (2014).
	\bibitem{global2} V. K. Chandrasekar, R. Gopal, A. Venkatesan, and M. Lakshmanan, \textit{Phys. Rev. E} {\bf 90}, 062913 (2014).
	\bibitem{global3} A. Mishra, C. Hens, M. Bose, P. K. Roy, and S. K. Dana, \textit{Phys. Rev. E} {\bf 92}, 062920 (2015).
	\bibitem{global4} G. C. Sethia, and A. Sen, \textit{Phys. Rev. Lett.} {\bf 112}, 144101 (2014).
	\bibitem{global5} F. B\"{o}hm, A. Zakharova, E. Sch\"{o}ll, and K. L\"{u}dge, \textit{Phys. Rev. E} {\bf 91}, 040901(R) (2015).
	\bibitem{global6} L. Schmidt, and K. Krischer, \textit{Phys. Rev. Lett.} {\bf 114}, 034101 (2015).
	\bibitem{global7} L. Schmidt, and K. Krischer, \textit{Chaos} {\bf 25}, 064401 (2015).	
	\bibitem{laing} C. R. Laing, \textit{Phys. Rev. E} {\bf 92}, 050904(R) (2015).
	\bibitem{hr_bera2} B. K. Bera, and D. Ghosh, \textit{Phys. Rev. E} {\bf 93}, 052223 (2016).
	\bibitem{local1} J. Hizanidis, N. Lazarides, and G. P. Tsironis, \textit{Phys. Rev. E} {\bf 94}, 032219 (2016).
	\bibitem{ch_hetero} J. Hizanidis, N. Lazarides, and G. P. Tsironis, \textit{Chaos} {\bf 19}, 013113 (2009).
	\bibitem{complex_ch1} Y. Zhu, Z. Zheng, and J. Yang, \textit{Phys. Rev. E} {\bf 89}, 022914 (2014).
	\bibitem{complex_ch2} A. Buscarino, M. Frasca, L. V. Gambuzza, and P. H\"{o}vel, \textit{Phys. Rev. E} {\bf 91}, 022817 (2015).
	\bibitem{multiplex1} S. Majhi, M. Perc, and D. Ghosh, \textit{Sci. Rep.} {\bf 6}, 39033 (2016).	
	\bibitem{chimera_multiplex} V. A. Maksimenko, V. V. Makarov, B. K. Bera, D. Ghosh, S. K. Dana, M. V. Goremyko, N. S. Frolov, A. A. Koronovskii, and A. E. Hramov, \textit{Phys. Rev. E} {\bf 94}, 052205 (2016).
	\bibitem{multiplex2} S. Ghosh, and S. Jalan, \textit{Int. J. Bifur. Chaos} {\bf 26}, 1650120 (2016).
	\bibitem{multiplex3} S. Ghosh, A. Kumar, A. Zakharova, and S. Jalan, \textit{Europhys. Letts.} {\bf 115}, 60005 (2016).
	\bibitem{multiplex4} S. Majhi, M. Perc, and D. Ghosh, \textit{Chaos} {\bf 27}, 073109 (2017).		
	\bibitem{two_nonlocal}  K. Premalatha, V. K. Chandrasekar, M. Senthilvelan, and M. Lakshmanan, \textit{Phys. Rev. E} {\bf 94}, 012311 (2016).
	\bibitem{two_global}  K. Premalatha, V. K. Chandrasekar, M. Senthilvelan, and M. Lakshmanan, \textit{Phys. Rev. E} {\bf 95}, 022208 (2017).	
	\bibitem{amc} G. C. Sethia, A. Sen, and G. L. Johnston, \textit{Phys. Rev. E} {\bf 88}, 042917 (2013).	
	\bibitem{cd_prl} A. Zakharova, M. Kapeller, and E. Sch\"{o}ll, \textit{Phys. Rev. Lett.} {\bf 112}, 154101 (2014).	
	\bibitem{imperfect_chi} T. Kapitaniak, P. Kuzma, J. Wojewoda, K. Czolczynski, and Y. Maistrenko, \textit{Sci. Rep.} {\bf 4}, 6379 (2014).
	\bibitem{travelling_chi} J. Xie, E. Knobloch, and H. C. Kao, \textit{Phys. Rev. E} {\bf 90}, 022919 (2014).
	\bibitem{hr_bera3} B. K. Bera, D. Ghosh, and T. Banerjee, \textit{Phys. Rev. E} {\bf 94}, 012215 (2016).
	\bibitem{breath1} D. M. Abrams, and R. Mirollo, S. H. Strogatz, and D. A. Wiley, \textit{Phys. Rev. Lett.} {\bf 101}, 084103 (2008).	
	\bibitem{spiral_chi} B. W. Li, and H. Dierckx, \textit{Phys. Rev. E} {\bf 93}, 020202(R) (2016).
	\bibitem{2d} J. Xie, E. Knobloch, and H.-C. Kao, \textit{Phys. Rev. E} {\bf 92}, 042921 (2015).
	\bibitem{3d} Y. Maistrenko, O. Sudakov, O. Osiv, and V. Maistrenko, \textit{ New J.	Phys.} {\bf 17}, 073037 (2015).
	\bibitem{pre2017} A. Schmidt, T. Kasimatis, J. Hizanidis, A. Provata, and  P. H\"{o}vel, \textit{Phys. Rev. E} {\bf 95}, 032224 (2017).	
	\bibitem{2D_linchi} C. H. Tian, X. Y. Zhang, Z. H. Wang, Z. H. Liu, \textit{Front. Phys.} {\bf 12}, 128904 (2017).	
	\bibitem{kura_book} Y. Kuramoto, Chemical Oscillations, Waves, and	Turbulence (Dover Publications, Inc., Mineola, NY,2003).
	\bibitem{nonlinear_coup} Consider the equation in complex form $\dot{z}=H(z) = \tilde{a}^2z - z|z|^2$, where $z=x+iy$ and $\tilde{a}$ is real constant. Its equivalent cartesian form is $$\dot{x}=\tilde{a}^2 x-x(x^2+y^2), ~~\dot{y}=\tilde{a}^2 y-y(x^2+y^2).$$ This system represents a limit cycle oscillator with radius $\tilde{a}$ and center at origin. The term $-H(z_{i, j})$ in Eq.\eqref{eq:1} of $\dot{z}_{i, j}$ has an influence of pulling $z_{i, j}$ toward the above limit cycle. But the other term $H(z_{{i-1}, j})$ pushes away from the limit cycle. So the positivity of $H(z_{{i-1}, j})- H(z_{i, j})$ implies that the pushing effect is more than the effect of pulling and finally may lead to synchronization for appropriate amount of coupling strength.	
	\bibitem{moving_pre} L. Janagal, P. Parmananda, \textit{Phys. Rev. E} {\bf 86}, 056213 (2012).
	\bibitem{nakao2015}  H. Nakao, \textit{Contemp. Phys.} {\bf 57}, 188 (2016).
	\bibitem{syn_book}  A. S. Pikovsky, M. Rosenblum, and J. Kurths, \textit{Synchronization: A Universal Concept in Nonlinear Sciences,} {\bf 12}, Cambridge University Press, Cambridge (2001).
	\bibitem{ott1} E. Ott and T. M. Antonsen, \textit{Chaos} {\bf 18}, 037113 (2008).
	\bibitem{ott2} E. Ott and T. M. Antonsen, \textit{Chaos} {\bf 19}, 023117 (2009).	
	\bibitem{ott3} G. Bordyugov, A. Pikovsky, and M. Rosenblum, \textit{Phys. Rev. E} {\bf 82}, 035205 (2010).
	\bibitem{ott4} S. A. Marvel, R. E. Mirollo, and S. H. Strogatz, \textit{Chaos} {\bf 19}, 043104 (2009).
	\bibitem{angular_frq} T. Pereira, M. S. Baptista, and J. Kurths, \textit{Europhys. Lett.} {\bf 77}, 40006 (2007).	
	\bibitem{rulkov1} N.F. Rulkov, \textit{Phys. Rev. Lett.} {\bf 86}, 183 (2001).	
	\bibitem{rulkov2} G. de Vries, \textit{Phys. Rev. E} {\bf 64}, 051914 (2001).		
	\bibitem{analytical_signal} P. Panter \textit{Modulation, Noise, and Spectral Analysis} (McGraw-Hill, New York, 1965).	
	\bibitem{gabor} D. Gabor, \textit{J. Inst. Electr. Eng.} {\bf 93}, 429 (1946).
			



\end{thebibliography}
\end{document}